# Plausibility as Failure: How LLMs and Humans Co-Construct Epistemic Error

Cláudia Vale Oliveira[1 [0009-0006-7443-4786]], Nelson Zagalo[2 [0000-0002-5478-0650]],
Filipe Silva[3 [0009-0007-7684-4786]], Anabela Brandão[4 [0009-0009-7996-9351]],
Syeda Faryal Hussain Khurrum[5 [0009-0000-6758-7844]], and Joaquim Santos[6 [0000-0001-5376-4774]]

DigiMedia, University of Aveiro, Aveiro, Portugal
[1] claudia.mvoliveira@ua.pt
[2] nzagalo@ua.pt
[3] fas@ua.pt
[4] anabela.brandao@ua.pt
[5] syeda.faryal.khurrum@ua.pt
[6] jnsantos@ua.pt

**Abstract.** Large language models (LLMs) are increasingly used as epistemic partners in everyday reasoning, yet their errors remain predominantly analyzed through predictive metrics rather than through their interpretive effects on human judgement. This study examines how different forms of epistemic failure emerge, are masked, and are tolerated in human–AI interaction, where failure is understood as a relational breakdown shaped by model-generated plausibility and human interpretive judgment. We conducted a three-round, multi-LLM evaluation using interdisciplinary tasks and progressively differentiated assessment frameworks to observe how evaluators interpret model responses across linguistic, epistemic, and credibility dimensions. Our findings show that LLM errors shift from predictive (factual inaccuracy, unstable reasoning) to hermeneutic forms, where linguistic fluency, structural coherence, and superficially plausible citations conceal deeper distortions of meaning. Evaluators frequently conflated criteria such as correctness, relevance, bias, groundedness, and consistency, indicating that human judgement collapses analytical distinctions into intuitive heuristics shaped by form and fluency. Across rounds, we observed a systematic verification burden and cognitive drift: as tasks became denser, evaluators increasingly relied on surface cues, allowing erroneous yet well-formed answers to pass as credible. These results suggest that error is not solely a property of model behavior but a co-constructed outcome of generative plausibility and human interpretive shortcuts. Understanding AI epistemic failure therefore requires reframing evaluation as a relational interpretive process, where the boundary between system failure and human miscalibration becomes porous. The study provides implications for LLM assessment, digital literacy, and the design of trustworthy human–AI communication.

## 1 Introduction

The production and accessibility of knowledge is being reshaped by the widespread adoption of artificial intelligence (AI) systems, particularly large language models (LLMs). Despite their unprecedented informational reach, challenges inherent to their generative capacity are emerging across different industries. Beyond technical, organizational, and ethical concerns such as data quality, resource management, privacy, and transparency (Rezaei, 2025), these technologies are simulating and influencing human behaviors such as judgment, reasoning, and decision-making (Gerlich, 2024). As such, more than the technical accuracy and performance of these systems, the challenge extends to the interaction complexity (Passerini et al., 2025) and to the trust that society places in them (Dang & Li, 2025).

The current capabilities of AI do not match the inflated public perception that often places these systems at, or above, human levels of intelligence (Dang & Li, 2025; Southern, 2025). Reasoning remains an intrinsic human quality that distances AI from the human intelligence level known as Artificial General Intelligence (AGI) (AAAI, 2025, 2025; Swigonski, 2025; Wong, 2025). Currently, models are unable to reason like humans, providing only a simulation of reasoning dependent on specific instructions in prompts and the knowledge base used to train them, making the simulation weak and sometimes erratic (Swigonski, 2025).

Despite these limitations, LLMs have been rapidly adopted, often without sufficient public debate or clear accountability mechanisms to address problems such as data reliability, misinformation (AAAI, 2025; Southern, 2025), and data privacy and security. Therefore, both the implementation of rigorous evaluation methodologies, employing approaches and metrics that support and shield AI systems, and the regulation and implementation of ethical guidelines that promote trust, are essential for integrating these tools into the community (AAAI, 2025).

This process can be enriched by studies that go beyond assessing systems accuracy through benchmarks, embodying human judgment (Miller & Tang, 2025). In fact, understanding the nature of errors demands interdisciplinary methods capable of observing how they emerge, how they are interpreted, and how they propagate. Therefore, rather than measuring accuracy in isolation, we investigate how users perceive, tolerate, judge, or normalize different kinds of AI answers through a human lens. By analyzing three rounds of multi-LLM evaluations across multidisciplinary tasks, we aim to expose the evolving ecology of error in its linguistic form, cognitive concealment, and evaluative bias. The central research question guiding this work is thus not *"How wrong are LLMs?"*, but instead *"What kinds of errors do we fail to see, and why?"* and *"What might be the implications of this human-AI dynamic for society?"*.

To this end, this paper is divided into seven sections. After providing context in the first introduction section, the second section presents the state of the art, defining and exploring LLM errors not only from a technical, predictive, and generative point of view, but also in terms of human influence on this process. After understanding how errors emerge and how performance can be evaluated, the third section presents the methodology, namely the objectives, methods used, including the construction of the dimensions and criteria for evaluating errors and each round. The details of the study are presented in the fourth section, followed by the results in the fifth section. The discussion of the results takes place in the sixth section, ending with the conclusions and limitations in the seventh and final section.

## 2    From predictive to hermeneutic error

Research on AI errors has been developed along technical accuracy and human reliability (Rong et al., 2024). In machine learning, error might be defined as a quantifiable deviation between predicted and observed values, where the goal is to minimize this deviation by optimizing model parameters. It represents noise, something to be removed, not interpreted. As such, it fills a negative space within the logic of optimization, meaning the less error, the better the model (Bian & Priyadarshi, 2024).

The notion of error expanded from isolated prediction mistakes to systemic interpretative gaps, in deep learning. Neural networks began to exhibit behaviors that resist tracing, producing high-performance outputs without transparent causal explanations. This opacity shifted the evaluation from measurement to interpretability, introducing epistemic uncertainty as a structural feature of AI systems. The central problem might no longer be prediction accuracy, but the ability to explain and justify model reasoning in human terms. This gives rise to the paradigm of interpretability and explainability in an effort to reconstruct human-understandable reasons for algorithmic output (Rong et al., 2024).

The emergence of generative AI (GenAI) has transformed the traditional understanding of error. In LLMs, error no longer stands from a misalignment between a predicted value and a defined, verifiable ground truth (Wang et al., 2026). Models are optimized primarily through objective functions, such as maximizing the probability of the next token (Du et al., 2025), that inherently reward linguistic coherence and plausibility over factual veracity. LLMs excel at generating outputs that are fluid, grammatically impeccable, and semantically convincing, even when the underlying content is factually incorrect, logically inconsistent (Anh-Hoang et al., 2025), or ethically questionable (Deng et al., 2025). This phenomenon has become central to understanding AI's epistemic limits.

## 2.1 Understanding error manifestation

Natural Language Processing (NLP) is a branch of AI and computer science that enables machines to comprehend and interpret human language in both spoken and written forms (Lan et al., 2025). Recent progress in NLP and LLMs has led to improved contextual and semantic understanding, resulting in more coherent and accurate responses (Y. Lee et al., 2024). However, models operate based on statistical inference, without any awareness or understanding, and are limited to reproducing the patterns that emerge from the data on which they were trained (Crawford, 2021; Hoffmann, 2022) or are accessible to them (Fan et al., 2024), undermining the reliability of their outputs.

Hallucinations are usually referred to as plausible but factually incorrect or unverifiable content (Banerjee et al., 2025; Huang et al., 2025; Z. Ji et al., 2023; Maynez et al., 2020). There are two ways to categorize hallucinations. Intrinsic, when the output contradicts the source data, or extrinsic, when the model introduces unsupported content (Z. Ji et al., 2023). Nevertheless, hallucinations may not necessarily be considered errors when the output cannot be supported nor contradicted by factual sources (Maynez et al., 2020) or when the information is wrong in the training data (Kalai et al., 2025).

Several inaccuracies might be considered factual errors when false or fabricated statements are produced (Maynez et al., 2020) such as incorrect dates, locations, or authors, leading to misinformation and a loss of user trust (Huang et al., 2025; Z. Ji et al., 2023). Referential errors might occur when the attributed sources, such as authors or digital identifiers, are fabricated or incorrectly attributed. On the other hand, semantic errors occur when the model misinterprets meaning or context, often producing loosely associated or irrelevant responses (Z. Ji et al., 2023), or logical, when invalid or inconsistent reasoning steps compromise internal coherence, leading to faulty conclusions. Also, contextual errors disrupt dialogue coherence by changing the topic or tone without justification. Instructional or prompt-misunderstanding errors occur when the model misinterprets task requirements, for example, by generating a narrative rather than a list, thereby reducing task utility. Inferential errors occur when valid premises lead to invalid conclusions (Huang et al., 2025), epistemic hallucinations, when speculation is presented as certainty (Ling et al., 2024), and completion-level errors, when token-level biases result in plausible but incorrect word choices (Huang et al., 2025).

Faithfulness and truthfulness are two categories particularly significant in recent analyses. Faithfulness errors occur when the generated text is not faithful to the input or context, adding unsupported facts or interpretations (Huang et al., 2025). Truthfulness errors reflect the repetition of widespread human misconceptions or falsehoods, showing how models can mimic systematic biases present in human discourse (S. Lin et al., 2022). Creative or imaginative hallucinations occur when fictional elements are introduced without signaling, blurring the boundary between fact and invention (M. Lee, 2023). Closely related, confabulation describes cases where the model presents plausible explanations or details to fill informational gaps, simulating reasoning without genuine understanding (Bender et al., 2021; Farquhar et al., 2024).

These manifestations are beginning to actively surface as consumable content, not as mere curious experiments. Recently, cross-sector studies have confirmed the extent and persistence of these challenges. These studies reveal that 45% of responses generated by AI contain at least one significant problem, with source errors responsible for 31% of all problems (Fletcher & Verckist, 2025). Although AI assistants have improved their stylistic quality, journalists have discovered that they systematically misquote or fabricate sources, mixing facts and opinions in fluent narratives that the public perceives as credible. As generative systems are becoming more common for policy-making or communicating with citizens, similar concerns (OECD, 2025) are arising in the government context, with the downside that results can't be traced back to evidence, which can undermine accountability and transparency. The problem reaches science (Purificato et al., 2025), where the automation of reasoning triggers hallucinations and knowledge production without human control or supervision . These phenomena often lead to misinformation, where confidently presented but inaccurate statements are disseminated, sometimes indistinguishable from authentic knowledge (S. Lin et al., 2022). Whether it is misrepresentation, hallucinations, or transparency issues, one common factor among these challenges is system interpretability.

## 2.2 Performance evaluation and trends

LLMs comparative evaluations typically assess reasoning, factual accuracy, contextual understanding, and ethical alignment. Models such as GPT-4, PaLM 2, Claude, and LLaMA 3 perform strongly on benchmarks such as MMLU and BIG-Bench, although performance varies across industries (Anil et al., 2023; Bubeck et al., 2023; OpenAI et al., 2024). GPT-4 excels in reasoning and code generation, PaLM 2 demonstrates superior multilingual capabilities, Claude emphasizes safety and interpretability. LLaMA 3 and Mistral have advanced open research by achieving competitive results with comparatively smaller parameter sizes (Ni et al., 2025). Nevertheless, the challenges remain when it comes to evaluation consistency, transparency, and achieving balance between model capability and ethical safeguards.

In a technical environment, there is a tendency to evaluate emerging technologies through automated systems. Although this may be the standard procedure for machines and software, for conversational AI it may be slightly different. Performative and quantitative metrics of efficiency, predictability, and scalability matter, but so does the qualitative and interpretative approach (Kalai et al., 2025). By training systems to perform well on a specific benchmark, such as BLEU, ROUGE or USR, other dimensions, such as interpretation or conceptual understanding, can be overlooked (Bandi et al., 2025).

While BLEU is focused on speed and precision, when applied to conversational AI, these predictive evaluation systems can sometimes fall short on user expectations (Papineni et al., 2002). Similarly, ROUGE automates summary evaluation by comparing system-generated outputs against sets of human-created reference summaries. Although multi-document summary tasks remain more challenging for the system to handle, it has shown strong correlation with human-assigned scores in a single document and short summaries (C.-Y. Lin, 2004). Likewise, USR automates dialog evaluation through a reference-free approach design to capture the complexity and diversity inherent to conversations. It employs several sub-metrics such as naturalness, interestingness, and knowledge usage, combining them into a single *"Overall Quality"* metric. Furthermore, its lack of need for supervision enhances its scalability while still achieving correlations with human judgement. However, its dependence on pretrained models raises concerns for potential bias, and its focus on open-domain chatbots limits its versatility for structured close-domain tasks (Mehri & Eskenazi, 2020).

Given that these systems are gradually being integrated into society and made accessible to everyone, the impact of their use must be studied. Guided by ambitions and differing opinions from major companies in the field such as OpenAI and others, Chollet created a test based on cognitive science to assess whether these systems will be capable of achieving a human level of reasoning: the Abstraction and Reasoning Corpus for AGI (ARC-AGI). Models failed this test with results below 32%, including scores close to zero. Only o3 achieved a result similar to humans with 87%. However, Chollet made updates to the tests that brought the models' results back to minimum percentages (Wong, 2025). In addition to ARC-AGI, there is a set of benchmarks designed to evaluate specific indicators such as GLUE, MMLU, MATH, GPQA, and HumanEval (AAAI, 2025). However, the percentages are minimal. These results may indicate that memorizing and cataloging information is not the same as understanding and applying it wisely.

Despite these results, the general public currently has access by default to GPT5 in its free version at most, which indicates that even if these systems evolve, there is no guarantee that the average consumer will obtain a response with the required level of thinking and performance (Wong, 2025). Furthermore, the problem with benchmarks is that machines are being used to test machines, leading to gaps in assessments that do not consider human "tact" and perception.

The evaluation of these systems in uncontrolled environments, from the average user's perspective, is scarce and presents its challenges. The models' ability to learn from user feedback makes evaluation even more complex, since after interacting and learning from their feedback, it contaminates subsequent results (AAAI, 2025). Additionally, although models provide a single answer, it generates thousands and then decides which one to present to the user based on criteria irrelevant to the user. This raises questions about energy consumption and funding used to provide a single answer (Wong, 2025). Given the progressive and prominent investment in these technologies, which are not expected to reach a higher level of autonomy and knowledge soon, the cost/benefit ratio of these systems is balanced, as they directly influence and exhaust the resources vital to the functioning of the planet (Swigonski, 2025).

## 3    Methodology

The persistence of error in GenAI poses a fundamental challenge to epistemic accountability for both human and machine systems. While research addresses hallucination mitigation, little empirical work has explored how AI error is perceived, normalized, and legitimized in evaluative practice.

Therefore, the purpose of this study is not only to identify potential errors made by LLMs, from factual consistency to reasoning, but also to understand how humans perceive, interpret, and react to these errors. These two approaches provide a way to document what types of errors occur and in what context, either by establishing different types of prompts or by tracking the cognitive process that shapes the evaluative judgment.

To achieve these aims, we designed a multi-chat and multi-round evaluation protocol involving anonymized LLMs and human evaluators. Across three temporal rounds, the study systematically varied both the epistemic level of task complexity and the evaluation instruments, from multi-criteria Likert scales to heuristic checklists. This design allowed for the observation of how error patterns and evaluative behavior evolve when the task demands a shift from high data coverage to low certainty, how evaluators experience fatigue or expectation drift, or if model updates change generative behavior over time.

By combining quantitative scoring data with qualitative observations from evaluators and analysts, this study establishes an interpretive corpus that exposes not only the linguistic form of error, but also its social interpretation. Additionally, this study relies on establishing evaluation dimensions and criteria, and three evaluation cycles. The dataset and analytical procedures were developed to maintain replicability, blinding, and temporal comparability while reflecting realistic use conditions for LLMs in professional and everyday settings. Ultimately, we hope to connect the findings from all phases to identify patterns of reasoning among the chatbots.

Focusing on human assessment, the intention is to discuss potential problems and challenges that may emerge while interacting with these systems and build an understanding of AI while identifying and raising awareness of potential sources of misinformation.

### 3.1    Dimensions and evaluation criteria

Given the scarcity of resources for evaluating LLMs (J. Lee & Hockenmaier, 2025) there was a need to combine the existing dimensions in the evidence with customized dimensions adapted to the scope of this evaluation. Thus, the dimensions and criteria established were inspired by different frameworks.

The framework for evaluating chatbot responses proposed by Google assesses criteria such as *Safety, Groundedness, Informativeness, Citation Accuracy, Helpfulness* and *Role Consistency,* or *Sensibleness, Specificity and Interestingness* to establish *Quality* (Thoppilan et al., 2022). Another framework, determines the quality of chatbots interviews through *Informativeness* and determines criteria that explore user experience and engagement as well as ethical feedback indexes (Han et al., 2021).

A third framework, used to evaluate chatbots in a healthcare context, presents several evaluation criteria, including *Up-to-dateness*, for example, which is relevant when comparing models without accessing the web. Especially because some models provided unsolicited references (Abbasian et al., 2024).

Another author determines the quality of conversations through four dimensions, namely, *Dialogue behavior*, *Language expression*, *Response content* and *Recommendation Items*. Among the criteria assigned to each dimension, it is possible to find *Coherence*, *Naturalness* and *Explainability*. At the *Recommendation Items*, for example, *Novelty* is introduced as an uncommon criteria that measures whether it is capable of generate something new, exploring creativity (Chen et al., 2025).

Another survey uses *Groundedness*, *Validity*, *Coherence* and *Utility*. The author *"provides a unified taxonomy for evaluation criteria, a comprehensive review on existing metrics and their implementation, and tackle transferability between different metrics"* (J. Lee & Hockenmaier, 2025). Supporting this statement, it was found that there are common criteria among these studies. However, they have been given different denominations while sharing the same purpose. Furthermore, it appears that there is no standard set of criteria or dimensions for evaluating LLMs, possibly due to the diversity of industries and the specific objectives of each one. Therefore, Appendix I, contains a reference table that establishes the connection between denominations with the same purpose for each criterion.

Once this redundancy had been resolved, we established the criteria and dimensions to be used in the first and second rounds, adapting them both to the study specifications and to the fact that they are instruments of human assessment. As a result, the following dimensions emerged: *Logical Reasoning*, *Dialogic Coherence*, *Utility*, *Language Expression and Complexity*, and *Ethical and Bias Challenges*. In addition, we introduced *Credibility and Citation* as a custom dimension aimed at identifying *Source Reliability*, *Existence*, and *Topic Relation*, which the research team deemed relevant after the Exploratory study.

## 3.2    Cycles of evaluation

The three-round design was not conceived as simple replication, but as progressive epistemic escalation. Each round deliberately changed both task structure and evaluative instrument, to trigger different generative failures and to capture the dynamic interaction between human judgment and machine feedback.

This design decision was grounded in the prediction that prolonged exposure to repetitive evaluation tasks can lead to fatigue and confusion, reducing the critical sensitivity of both evaluators and analysts. By varying instruments and task typology, subsequent rounds sought epistemic differentiation, avoiding redundancy and interpretive misjudgments. Conducting the rounds over separate months also allowed us to observe potential model and evaluation drift, revealing changes in both AI behavior and human calibration over time.

## 3.3    Assessment scales

The trial-and-error process in the exploratory study considered different evaluation scales, such as SimpleQA, which uses the labels *Correct*, *Incorrect*, and *Not attempted* (OpenAI, 2024) and multi-level Likert Scales employed in previous studies by T. Ji et al. (2022), Abeysinghe & Circi (2024), and Chiu & Chung (2024).

Cross-referencing the findings obtained from the exploratory study against the methods used within the framework proposed by Tam et al. (2024), identified in Appendix I, Likert scales and binary scales such as *Presence/Absence* are emphasized. However, binary scales provide a very straightforward assessment when compared to the complexity of the responses provided by LLMs, as well as the complexity of the human analytical process. For example, if three quotes are provided in a conversation and one of them is unrelated to the topic, how can it be properly evaluated? As such, a 5-point Likert scale was selected to evaluate each criterion, where 1 represents *Never*, 2 *A few times*, 3 *Sometimes*, 4 *Almost always* and 5 *Always*.

This option was used in the first and second rounds, whereas in the third round a binary scale (Yes/No) is used, complemented by an observations field.

# 4 Presentation of the study

The present study is formed by two stages, starting with an exploratory experiment followed by the main experiment divided into three rounds.

## 4.1 Exploratory study

For the exploratory study, conducted in February 2025, several chatbots were selected, namely ChatGPT, Claude, CoPilot, DeepSeek, Gemini and Perplexity. The goal was to test free available models by submitting a series of questions on general knowledge, both national and international, asking about universal truths, mathematical problems, ethical dilemmas, biased questions, and riddles, as well as preparing academic summaries and a comprehensive analysis of tables. In the first phase, each researcher conducted their own test with customized subjects, interacting freely with each chatbot in both Portuguese and English languages. In the second phase, sets of sequential questions were defined in English to identify and understand potential variations provided by each chatbot in similar conditions.

Although it was not possible to justify the results with concrete data extracted from the functioning of each chat, some problems were identified regarding cited sources, logic, reasoning, entailment, moral values, the interpretation of provided documents, and cultural evidence from small countries such as Portugal. The feedback obtained was not direct in the sense of "black and white" answers, even when the question requires it. There is a tendency to always provide an answer, even if the content is not appropriate or precise, thus increasing noise. It seems it tries to either clarify with similar subjects or provide anchors to continue the conversation, creating a vicious loop.

After discussing the results, the chatbots to be tested in the main study were chosen, namely the ChatGPT (version 4o), Gemini (version 2.5-pro), DeepSeek (version r1-0528-maas) and LeChat (version Mistral Medium 3). This selection is based both on benchmarking indexes (Artificial Analysis, 2025), consulted in March 2025, and on the attempt to achieve greater representativeness, opting for the two most dominant Americans, one Chinese, and one European. This need for greater representation also arose after noticing cultural deficits in answers, especially when comparing prompt about small countries with world leaders.

## 4.2 Round design

In the first round (R1), we adopted an evaluation strategy based on the Dimensions and evaluation criteria, and specific questions associated with each criterion (see Appendix II). The evaluator's focus would be to understand the criterion through the question and evaluate the chatbot feedback in line with their understanding, providing a *Score* based on the Likert scale established in the chapter 3.3, an *Explanation* for the assignment of that score, and record anomalies or other observations in the field *Other findings*.

In this round, the qualitative value was emphasized over the quantitative value. It became clear that the problem was not just about the quality of the chatbot's response itself, but rather about the expectations and needs of the user, who express dissatisfaction even when an answer was factually correct. As a result, the strategy for the second round (R2) was redesigned introducing criteria that demonstrate user satisfaction, namely *Clarity* and *Satisfaction* (Tam et al., 2024). To reduce the cognitive load on evaluators, the dimensions and criteria columns were removed from the evaluation form, retaining only the questions about them (see Appendix III). In addition, some questions were reworded in an attempt to increase the interpretive clarity of the nuances of each criterion, even if it is not directly identified on the form. By consulting Appendix IV, it is possible to preview the new criteria and dimensions, available only to analysts, and the corresponding questions, available only to evaluators.

It was found that solutions to riddles and puzzles were already embedded in the models training. This generates a biased cycle where the system only transmits information rather than providing autonomous reasoning similar to humans, making it impossible for new or novelty information to emerge (Wong, 2025). Therefore, in R2, the prompts included puzzles or problems that are unknown or non-standardized, meaning that there is no direct answer that can be found through web search, limiting the probability of them being already used to train models or appearing as a source of citation.

The approaches used in the first rounds (R1/R2), while rigorous, proved to be demanding and difficult, especially while doing systematic and prolonged evaluation. As a result, in addition to gradually reducing the number of prompts to be evaluated, in the third round (R3), we opted for a more heuristic and functional approach, focused on directly detecting errors, falsehoods, and problems (see Appendix V). This transition was driven by three factors. The first consists of improving the efficiency of the process by reducing the time and effort required by evaluators and analysts, while maintaining the ability to detect problems without influencing participants. On the other hand, this approach is closer to real world action, where determining whether information is reliable or doubtful is a fluid process that depends on the independent cognitive ability of each person. By simplifying this process, we shift the focus of detailed analysis to the researcher, who can freely interpret and identify patterns.

Transversely across all rounds, some specific adjustments were made to the instructions given to the involved research team (managers, evaluators, and analysts). An overview of the rounds is available in Table 1 where it is possible to identify the selected chatbots, the type of question and explanation requested in the evaluation forms, and the number and type of prompts that were evaluated.

| Experiment | Chatbots | Question type | Explanation type | Prompts Quantity x Type |
|---|---|---|---|---|
| Round1 (R1) | | Dimension/Criteria lens | Score, Explanation and Other findings | 10 x Real events ambiguity |
| Round2 (R2) | ChatGPT, Gemini, Claude, CoPilot | Overall satisfaction and performance evaluation | Explanation, Score | 7 x Everyday problems |
| Round3 (R3) | | Errors report | Yes/No, Description | 6 x Error injunction |

Table 1. Experiments overview

Regarding prompt typology, R1 included questions related to specific industries and events, namely the Olympic Games in the sports category, the Grammys in the music category, the Oscars for cinema, activism in the arts, and the best game of the year in the technology category. The questions were designed to activate critical processing, for example, *"What public health problem stained the 2024 Olympic Games?"* and *"What was the main debate about mental health and violence that emerged from the 2022 Oscars ceremony?"*. These are questions with predictable answers, but they do not have a unique answer. They might trigger open-ended answers and reasoning that deviate from the expected.

On the other hand, the questions designed for R2 were more restricted with no answers to be expected, especially those related to everyday problems such as the blackout on April 28, 2025 (in Portugal and Spain) or clarity on passenger rights on public transportation. For R3, the questions were specifically designed to mislead in an attempt to map flaws, both in the answers provided by the chatbots and in the human ability to detect them, using questions based on non-existent elements, such as unknown charts, entities or events.

## 4.3 The evaluation process

The evaluation process was structured around strict role separation and temporal sequencing to ensure both validity and interpretive richness.

### Participants, Roles and Blinding

All team members are academic researchers in the multidisciplinary field of communication technologies and sciences. The group consists of four doctoral candidates and two professors, who together bring a combined experience of over 30 years in teaching and research. They are currently exploring the role and impact of generative AI within their respective fields of research, such as Human-Computer Interaction, Storytelling processes, Creativity processes, VR/AR development, Literacy in Higher Education, and Secondary Teaching. Ranging from early-career scholars to established academics, their participation provides informed data on the implications of GenAI.

Researchers were divided in three participant profiles in the study. The manager (1) oversaw all procedural steps, anonymized model outputs, and distributed materials to evaluators and analysts without revealing system identities. The evaluators (3) scored or annotated each AI response according to the instrument provided for each round, offering optional free-text remarks explaining their reasoning. After the evaluators, the analysts (2) performed the same procedure and only then reviewed the evaluators' feedback, identifying recurrent patterns of interpretation, disagreement, and misclassification of error types.

No evaluator had access to other participants' judgments or to the original model prompts beyond the assigned set. This design maintained double blinding, which means, evaluators were blind to the model, and analysts were blind to the identity of evaluators and models.

### Materials and Instruments

Each round employed a dedicated set of textual instruments, such as instructions, prompt sheets and evaluation forms. *Manager Instructions* detailing the blinding and anonymization procedures. *Evaluator Instructions* specifying scoring criteria, examples (in some cases), and expected reflection length. *Prompt Sheets* containing the task items and the *evaluation forms* captured numeric ratings and open remarks.

All materials were standardized in format and language, ensuring comparability across rounds. The resulting corpus (prompts, responses, evaluations, remarks) formed a layered dataset suitable for both content and discourse analysis.

### Data collection, procedures and limitations

Prior to prompting, the manager verified and, when necessary, activated the web-browsing capabilities of each chatbot. Although the initial intention was to evaluate the models without browsing enabled, this configuration was unattainable for *Gemini*, as the browsing function remained permanently activated with no option to disable it.

The manager prompted the chatbots using the question set previously established by the analysts. After collecting the outputs, the manager saved each response in a separate document and proceeded to code both the chatbots and the three human evaluators.

More specifically, upon receiving the prompts, the manager accessed the web-based chatbots through the Opera browser's built-in VPN, configured to a European location. This procedure, however, was not feasible for all models. *Gemini* (Alphabet/Google) did not permit access via VPN. Similarly, although the manager aimed to interact without logging to avoid potential challenges related to user history or cache, this was again not possible with *Gemini*. In this case, a throwaway Gmail account was created solely for the purpose of generating the responses.

Each question was submitted within a new, isolated chatbot session. The resulting output was then transferred into the template supplied by the analysts, including shareable conversation links when available. This procedure was repeated for all chatbots.

The manager was responsible for coding both the AI agents and the evaluators, thereby ensuring that the analysts remained blind to the identity of both groups throughout the process. Therefore, he subsequently organized the files into a dedicated folder, naming each prompt sheets with the variable Cx, where x stands for the chatbot code (1–4), and Ex for each evaluation forms, where x stands for the evaluator number (1–3).

This protocol was repeated across all three rounds of data collection.

**Analytical procedures and limitations**

Across all rounds, analysts performed the same assessment process as evaluators before assuming their role as analysts, since there is a *"tendency to perceive events, once they have occurred, as more predictable than they actually were"* (Passerini et al., 2025). An *a posteriori* evaluation of these results could lead to biased findings. Therefore, to avoid this possibility, the analysts evaluated the feedback from the prompts before interpreting the feedback from the evaluators.

In R1, analysts highlighted cases where evaluators gave scores different from 1 or 5 and/or where they added comments. However, some coincidences were found in the scores that challenge this approach, raising questions about whether it would be possible to find other patterns by performing a more thorough analysis. Besides, there was an incident when pasting the citations into the materials provided to the evaluators and analysts. As such, the *Credibility and Citations* dimension was not considered in this round, affecting the evaluation of these criteria.

In R2, all cases were analyzed regardless of the score assigned by the evaluators, unlike in the first round. All chatbot responses and references provided were also analyzed to compare with the feedback, or absence of feedback, from the evaluators. Also, a comparison was made between the observations registered at the same prompt across all chats, just like in R3. It was a thorough but necessary process to identify the details that shape the results of this analysis.

# 5    Analysis results

The detailed results of each round, specifying each prompt, chat, and evaluator (if applicable), can be found in the Appendix VI. For clarity, this chapter is organized by three layers. In section 5.2 recurrent error patterns in chatbot behavior based on the comparative analysis of outputs across all rounds are identified. In section 5.1 error patterns in evaluators' cognitive behavior are examined, focusing on how criteria were operationalized, confused, or collapsed in practice. The porous zone between these domains is addressed in section 5.3, where epistemic failure becomes co-constructed through interaction and cannot be attributed solely to either the models or the humans.

## 5.1    Chatbots performance patterns

Among all four evaluated chats, some patterns of performance and behavior emerge. DeepSeek tends to produce exhaustive and dense answers, although it often digresses and presents redundancies and vague information supported by outdated, non-existent, or irrelevant sources. On several occasions, DeepSeek presented internal and reasoning contradictions or prompt misinterpretations. Examples include the answers DeepSeek gave to fourth and seventh prompts in R2. In the first (R2/DP/P4), when asked about changes made to the National Health Service (SNS) in March 2025, the chatbot confused healthcare systems by presenting options for Portugal, Spain, and Australia. This blend of healthcare systems from different countries could be considered inconsistent.

In the second (R2/DP/P7), DeepSeek was provided with a meeting schedule and the time it took to get to the departure train station and then from the arrival station to the meeting location. However, in addition to not providing references to justify the choice of the best bus schedule to arrive on time, it provided extra time at the start of the trip to arrive on time at the departure terminal but did not consider this time margin to arrive slightly before the meeting at the destination.

ChatGPT presented redundancies and contradictions in both answers and sources in terms of consistency, reasoning, and correctness. One example was when it mentioned the commitment of doctors, stating that *"the introduction of a total commitment regime for physicians, initially voluntary and later mandatory for management positions, aims to increase the availability and commitment of professionals"*, when the fourth prompt of R2 (R2/GPT/P4) explicitly asks how changes to SNS status affect patients. However, despite claiming that it "directly benefits patients," it is not clear how this measure affects them, not only because it is vague, but mainly because it is voluntary.

Moreover, the source used to support this argument is not only outdated, being an article from 2021, but also refers only to measures applied to the management and executive leadership of the SNS itself. This case is one of the examples were ChatGPT presented outdated references that are sometimes fabricated, non-existent or unrelated to the topic. When asked about the causes of rising crime in Portugal, in R2, second prompt (R2/GPT/P2), the chatbot presented arguments based on data collected up to 2023. On the other hand, in R3, when replying to the second prompt questioning laws restricting the use of AI in public schools (R3/GPT/P2), the chatbot provided a series of unreliable links such as Reddit posts or non-existent pages.

Gemini revealed a tendency to offer long, structured answers but not always relevant or related to the subject. In fact, when responding to the third prompt in R2 (R2/GM/P2), which asks about bus passengers' rights when a trip is delayed, Gemini provided 1,551 words. A huge difference compared to LeChat with 386 words, DeepSeek with 910, and ChatGPT with 914. However, the quality of the information is not proportional to the number of words. The textual density and marked presence of references can create an illusion of validity.

This represents a concerning finding, given that Gemini confirmed the existence of a non-existent event (R3/GM/P6), the AI & Democracy Summit held in Brussels in April 2025. It started by describing the event and provided a list of outcomes and topics discussed such as *"Strengthening AI Governance and Regulation", "Addressing Risks to Democracy"* and *"Leveraging AI for Democratic Enhancement".* In addition, it supported the arguments with credible references, but not related to the event itself, only to topics that the event could have if it existed.

Also, Gemini showed indicators of bias reproduction when stating, in answer to the sixth prompt of R2 (R2/GM/P6), that in Portugal there is a community with stereotypical beliefs and prejudice. When asked about the social obstacles of emigrating to Portugal as a Brazilian, the chatbot said that *"although generally welcoming, some negative stereotypes or prejudices towards immigrants can exist"* based solely on a blog entry.

Ultimately, Gemini accurately states that the UNESCO Charter on AI Education mentioned in the first prompt of R3 (R3/GM/P1) does not exist but provides a list of changes that have been made by UNESCO in this field. After analyzing, if the charter does not exist, there are no changes to list because there is no comparison between the new charter and the current one. Everything the chatbot presents after that is speculation based on documents it finds constituting misleading information.

Similarly to Gemini, LeChat, in the first prompt of R3 (R3/LC/P1), also assumed the existence of the UNESCO Charter with a clear statement supported by inadequate references documents, citing:

> *"The UNESCO Charter on AI Education, adopted in May 2025, introduced several key changes aimed at integrating artificial intelligence into educational systems responsibly and effectively. (...) These changes reflect UNESCO's commitment to leveraging AI to improve educational outcomes while addressing the ethical, social, and practical challenges associated with its implementation".*

It tends to provide irrelevant or misplaced information, often introducing unsolicited content. The tenth prompt from R1 (R1/LC/P10) demanded the planning of a trip to Japan in the spring, mentioning a desire to see the cherry blossoms, and introduced a change in dates, stating that the trip could only take place in mid-May. However, LeChat provided a series of suggestions such as events that were taking place outside of that season, such as three Cherry Blossom Festivals, and only at the end of the answers it mentioned that *"By mid-May, the cherry blossoms have usually finished blooming in most areas"* and that it will only be possible to see them if the travel dates are flexible.

On the other hand, LeChat, in the third prompt of R2 (R2/LC/P3) provided an answer based on flight regulations and policies, when the prompt explicitly requested information about bus refunds. In addition, it stated that full refunds could be accepted if requested before departure, which is not true, and presented refund options that were not related to travel delays, a requirement defined in the prompt. Also, the transport company number provided was incorrect.

Overall, the results indicate that all chatbots share a fragile semantic comprehension and logical reasoning, resorting to a combination of keywords to construct plausible, but not necessarily correct, answers. Across the three rounds, the four chatbots exhibited recurrent patterns of error that go beyond simple factual inaccuracy. These patterns concern how the models construct coherence, handle context, and simulate grounding. Rather than appearing as isolated mistakes, they form families of behavior that recurred across prompts, rounds, and systems, such as, semantic drift, internal inconsistency, contextual misalignment, misplaced information, false assumptions, referential fabrication, week grounding, speculative completion and overextended answers.

## 5.2    Evaluator behavior patterns

To understand the ecology of error, this section examines how evaluators interpreted, operationalized, and applied the assessment criteria across rounds, and how their judgement shifted over time.

Regarding the patterns found in the evaluators' assessments, a literal reading of the answers prevailed in R1. Explanations were often superficial and, in some cases, contradictory when cross-referenced with the scores assigned. In certain cases, they gave positive evaluations along with negative comments and sometimes provided no explanation for the scores they had assigned. Also, evaluators felt the need to obtain comprehensive answers, stating that *"the data is not fully justified as the response is too brief"* even though the chatbot presented a clear and complete answer. In addition, they yearn for details beyond what is requested in the prompt. An example is when the prompt asks for only one problem in a specific context and the evaluator is aware of several: *"the query says what public health problem stained the Olympics but there were several. The AI agent only lists one*." Evaluators rarely reported irrelevant or misplaced information, as in R1/LC/P10, suggesting that the way information is presented is more decisive than its contextual relevance. The inclusion of additional or irrelevant details was perceived as adding value across rounds.

In R2, there is greater attention to references, but without a coherent or consistent method for evaluating their quality, such as in R2/GM/P2 where the illusion of validity was not always criticized/reported by evaluators. Associating credibility with the number of sources rather than their authenticity or relevance seems to be the trend. The evaluators' perceptions are not consistent. At times, having few references negatively affects credibility, just as having many references increases satisfaction. In fact, credibility and clarity were still highly scored on several occasions, which indicates an overreliance on the formal appearance and density of answers and sources.

However, the way these references are assessed is not methodical. Sometimes the sources are credible, yet evaluators felt the need to seek information elsewhere from other sources they considered more relevant. Other times, the sources are not credible or contain accessibility errors, and yet the chatbot is still rated positively. One example is described in R2/DP/P7, where mostly positive evaluations were attributed, demonstrating that evaluators valued apparent effort and text length more than factual accuracy. In this round, providing several references increased credibility, just as the density of the answer increased the satisfaction of the evaluators. Also, participants rarely agree when evaluating the most controversial prompts identified in these experiments. One example is the evaluation of *Relevance* at the seventh prompt in R2 about the travel schedule. One evaluator states that the chatbot *"does not provide an actual train schedule"* assigning it the minimum score, other states that *"there is no train timetable or travel costs"* assigning it the average score, and the last one makes no comment and assigns the maximum score.

They seem to have different evaluation methods while evaluating similar situations. This statement refers to the experience analysis regarding the information's credibility, where a duality of expectations can be identified towards the references' authenticity, showing changes in the requirements for evaluating the same criterion. For example, in Gemini's responses to R2/GM/P6, or to R2/GM/P5, the references provided are mediocre, yet the score given by the evaluators is positive. On the other hand, there are situations in which references are credible, the information is factually correct, and the evaluations/comments are negative.

Throughout this first two rounds, the evaluators' progress shows a gradual familiarization with the criteria, but not necessarily an improvement in their evaluative accuracy. In R1, the inclusion of multiple criteria (*entailment, correctness, depth, bias*, etc.) seems to have caused some disruption in the evaluation, leading to confusion between dimensions, for example, *Correctness* being confused with *Depth* or *Up-to-dateness*, and *Relevance* being confused with *Agreement* or *Consistency*. Evaluators often confuse different criteria, making it challenging to quantitatively assess chatbot performance. Although the criteria in R2 are provided only as questions, confusion between dimensions persisted.

Evaluators collapsed criteria through intuition. Table 2 maps how original criteria are reinterpreted in R1 and R2. These distortions indicate that, although the instrument offered a fine-grained grid of dimensions, in practice evaluators compressed them into a handful of global impressions about the answer's completeness, clarity, and stylistic naturalness. Furthermore, some cases were identified where evaluators did not understand the answer provided by the chats, or there was a discrepancy between the score and the explanation provided by them, suggesting that the final score is driven by an overall "feeling" rather than by systematic application of the individual criterion.

| Original Criterion | R1 Distortion | R2 Distortion |
|---|---|---|
| Entailment | Correctness; Depth; Disambiguation; | N/A |
| Correctness | Depth; Bias; Consistency; | Up-to-dateness |
| Consistency | Agreement; Entailment; | N/A |
| Agreement | Depth; Bias; | N/A |
| Depth | Bias; Agreement; Disambiguation; Correctness; Naturalness | Comprehensiveness |
| Relevance | Agreement; Depth; Bias; Consistency; Entailment | N/A |
| Understanding | Naturalness | N/A |
| Bias | Relevance | N/A |
| Toxicity | Groundedness | N/A |

| Groundedness | Depth | N/A |
|---|---|---|
| Up-to-dateness | Depth; Disambiguation | Groundedness |
| Usefulness | N/A | Topic relation or General Credibility |
| Comprehensiveness | N/A | Agreement |
| Topic Relation | N/A | Reliability |

Table 2. Criteria distortion map identified in R1 and R2

R3 introduced a more analytical approach, but errors of interpretation persisted. Differences between evaluators remained, revealing the subjectivity of judgment. Even consistent evaluators (such as A3, who frequently identified biases and inconsistencies) continued to diverge from the others in their interpretation of semantic nuances. This confirmed that, even when the task is simplified, the evaluation of LLM outputs remains highly sensitive to individual calibration, tolerance for ambiguity, and personal thresholds for what counts as good enough. Evaluators provided different explanations when their scores are unanimous. This happened in R3, sixth prompt (R3/LC/P6), for example, when LeChat said that there is no law forbidding AI in public schools and mentions an existing Portuguese measure. Even so, each evaluator provided a different point of view for the same score: that the information is not up to date (despite the measure should be implemented by 2030), that the answer is incomplete (perhaps because it provides the information succinctly) or that the answer is not related to the topic.

Across the three rounds, evaluators demonstrated a tendency to value the answers' density and the number of references as indicators of quality, regardless of their credibility or relevance. Some examples are R1/LC/P10, R2/DP/P3, and R3/GPT/P2. In general, all evaluators demonstrated a need for more comprehensive and explanatory feedback beyond what is requested in the prompt and provided by the answer. The *depth* of the answer, as well as the quantity of references, significantly influenced the evaluators' satisfaction, even if credibility and quality were questionable.

This correlation between depth and the evaluator satisfaction confirms a perception of discursive validity, i.e., the more extensive and formally structured the response, the greater the sense of completeness and reliability. However, the density of information masks internal inconsistencies, contradictions, and even unnoticed factual errors. This may reveal an implicit expectation of proactive assistance from the chatbot, i.e., they expect the AI to go beyond the question, providing additional context, hypotheses, and comments, even if it was not requested.

In sum, three categories of patterns are revealed: (i) reliance on surface cues such as length and number of references, (ii) collapse of multiple criteria into a few intuitive macro-judgements, and (iii) evaluation drift and disagreements across rounds. These patterns confirmed that humans are not a neutral detector of machine error. Evaluators brought their own heuristics, compressed complex instruments into simpler impressions, and sometimes rewarded the features that allow unnoticed errors to pass, such as fluency, length, and apparent completeness. The next subsection examines how model behavior and evaluator cognition intersect to form a porous zone where hermeneutic error is co-constructed with the user rather than simply produced by the machine.

### 5.3    The porous zone, a hermeneutic co-construction of errors

The error categories and patterns described in 5.1 and 5.2 rarely occurred in isolation. Across rounds, the most problematic cases emerged in what this study termed as a porous zone between model output and human evaluation. In this zone, epistemic failure is not simply produced by the system and then detected (or not) by evaluators. It is gradually shaped as meaning and negotiation occur between a plausible answer and a cognitively overloaded reader.

Three recurrent mechanisms characterize this porous zone. First, there is an alignment between model plausibility and human surface heuristics. Fluent, well-structured responses invite trust, which makes evaluators more likely to repair and fill in the gaps, resolve contradictions, or treat speculative claims as reasonable elaborations. Evaluators demonstrated difficulty in detecting logical contradictions and subtle biases. Even when the content is biased, contradictory, or outdated, the scores remain mostly positive. This suggests that the evaluation is strongly influenced by linguistic fluency and naturalness of response, rather than factual accuracy or argumentative consistency.

Second, both models and evaluators rely heavily on keyword matching or skimming methods instead of analytical checking. Shared attention to salient terms, for example, country names or institutional acronyms, can give the impression that an answer is referring to the right entities and events, even when the underlying connections are incorrect. Third, evaluation tends to drift towards a broad sense of informativeness or relevance. Once a response feels globally useful or "on topic", explicit verification of factual details, temporal consistency, or causal links becomes less likely.

Within this porous zone, the boundary between minor imprecision and substantive error is continuously renegotiated. Rather than being a stable property of model behavior, error appears as a hermeneutic event, a joint outcome of generative plausibility and human interpretive shortcuts. The expectation operates through normalization and tolerance for speculation. When models were unable to provide accurate information, they often offered alternatives such as related events, similar policies, or generic procedures, rather than signaling non-knowledge. Evaluators rarely reported or penalized this behavior. Their behavior oscillates between a subjective and generalist approach, based more on the discourse appearance of completeness and clarity than on the answer's objectivity.

By mapping these behavioral patterns, it was possible to identify recurrent heuristics, confusions between criteria, and forms of drift that shaped the evaluation process and contributed to the emergence of error. The analysis of each round reveals that the ecology of error does not emerge only from the models' behavior. It is co-produced by the way human evaluators read, interpret, and apply the criteria. Their scores, comments, and omissions reveal recurrent cognitive patterns that systematically shape how errors are perceived, amplified, or neutralized.

## 6    Discussion

The root causes of LLM hallucinations are multifactorial. First, the architecture and training objective of LLMs, predicting the next token, are optimized for linguistic coherence rather than factual accuracy (Bender et al., 2021; Lee, 2023). Second, the training data are heterogeneous and frequently noisy, including biased or false information collected from the web (Z. Ji et al., 2023). Third, prompting techniques can exacerbate or mitigate errors. Ambiguous or leading prompts may trigger hallucinations, while well-structured prompts reduce uncertainty (Velásquez-Henao & Cadavid-Higuita, 2023). However, the average user is not trained to know how to structure prompts, or interacting socially with these systems, as they are with humans (Weinstein et al., 2025). Finally, contextual limitations, such as finite context windows and lack of persistent memory, prevent models from maintaining factual consistency across extended discourse (Liao et al., 2023).

However, evaluating feedback from these systems using automated benchmarks may not accurately reflect their performance level against users' preferences, expectations, and context. Although human evaluations constitute the smallest part of studies (Rong et al., 2024), adopting qualitative instruments, informed by human judgment, may be relevant for evaluating LLMs and conversational agents (Z. Ji et al., 2023).

The three-round evaluation process produced a rich corpus of generative outputs, human assessments, and interpretive remarks that revealed how both AI fallibility and human judgment evolve under different epistemic demands. The results confirmed that errors change depending on the chatbot and prompt typology, while evaluators' ability to recognize it decreases as plausibility and linguistic fluency of the answers increase. A common factor across all rounds is that the more language resembles understanding, the harder it becomes to identify its mistakes.

In fact, users tend to overestimate the reliability of LLM answers. Using expressions that communicate certainty, such as *"I am sure..."* significantly increases the level of trust that users place in conversations. The same applies to longer answers, even when they are less accurate. The presentation style, the language used, and the length of the explanation have an influence on user confidence, regardless of the answer's quality (Steyvers et al., 2025).

In fact, human evaluation frequently relies on intuitive global judgments that substitute for finer analytical distinctions, a dynamic well established in research on heuristic reasoning (Kahneman, 2011; Tversky & Kahneman, 1974), cognitive miser theory (Fiske & Taylor, 1991), and heuristic processing in persuasion (Chaiken, 1980). The cognitive mechanism is simple. When an answer *"sounds right",* users assume it is right. Humans frequently rely on surface-level cues, such as tone, fluency, length, as heuristic shortcuts for expertise, a phenomenon long documented in cognitive psychology (Alter & Oppenheimer, 2009; Chaiken, 1980; Petty & Cacioppo, 1986). Under time pressure or cognitive load, users rely on these heuristic cues rather than analytical verification (Gilbert, 1991; Kahneman, 2011).

As such, it is common to project structure and comprehension where none exists, leading to a transparency illusion in which people believe they understand the system's reasoning, and that the system reciprocally understands them (Rozenblit & Keil, 2002; Waytz et al., 2010). Therefore, the way information is presented influences user trust and validation ability. This ability is also influenced by the impression of talking to a human, which is enhanced by fluent and articulate speech. Actually, users are more likely to be misled not by factual consistency, but by linguistic coherence, which influences their perception towards the structure and style that enhance their sense of trust. The result might be an amplification loop where plausible text is mistaken for reliable knowledge (Miller & Tang, 2025).

Addressing this phenomenon requires reframing errors not only as malfunctions to be eliminated, but as a structural feature of generative systems, a hermeneutic condition that reveals how meaning and truth are co-constructed between humans and machines. Errors do not arise only from miscalculations against an objective reality, but from the way these systems generate language by predicting patterns rather than verifying factual content. They are not a deviation from a ground truth, but a by-product of probabilities and expectations (Passerini et al., 2025).

This introduces a hermeneutic fallibility, in which the distinction between truth and plausibility collapses at the level of expression. A model may produce a sentence that is grammatically perfect, semantically rich, and contextually appropriate, yet factually false. More importantly, it may do it with persuasive confidence, just as in R3/LC/P1. The resulting hallucinations are not random failures, but plausibility artefacts, symptoms of a system that has mastered the linguistic form while lacking referential verification.

The consequences of these errors extend beyond technical performance, and eliminating them from the systems may be complicated (Banerjee et al., 2025). At a social level, recurrent inaccuracies might erode human trust in AI systems (S. Lin et al., 2022). Misleading outputs affect interpretation and decision-making (Gill et al., 2024), threaten credibility, accountability, and the integrity of information dissemination (Nanz et al., 2025; Rodrigues, 2020). As a result, LLM errors have implications not only for systems design but also for epistemic responsibility in human–AI interaction.

Apart from trust being a subjective behavior dependent on human expectations and perceptions, its assessment may be unbalanced as it is influenced by the degree of transparency provided by algorithms and data sources (Heersmink et al., 2024). Trust strongly influences the safety and efficiency of human–AI collaboration, as both overtrust and undertrust pose risks (Afroogh et al., 2024). However, users tend to trust processes and content from automated systems, even if they contain errors. In fact, the results revealed a cycle of mutual reinforcement between chatbot rhetoric and evaluators' perceptions, showing that the more convincing the discourse, the greater the acritical content acceptance. One example is the fact that none of the evaluators reported what happened in R3/GM/P1, accepting the answer provided by Gemini, even though it was speculative, since the charter did not exist. This phenomenon is consistent with previous studies on *automation bias*, in which users tend to trust the apparent authority of automated systems, even when faced with obvious errors (Passerini et al., 2025; Romeo & Conti, 2025).

Therefore, both chatbots and users operate within margins of ambiguity. The former produce plausible but logically fragile responses. The latter evaluate based on perceptions. This phenomenon creates challenges for the critical analysis and validation of answers, leading to situations where the provided content is used blindly. Cases such as the controversy surrounding the hallucinations found in the Deloitte report (Paoli, 2025), are examples of how factually incorrect and fabricated content is being used without rigorous human filtering and validation. Consequently, it is crucial to raise awareness among users regarding the responsible use of these tools, warning them about the fact that fluency and quantity are not equivalent to accuracy and quality.

This misalignment between apparent and actual quality is the main transversal pattern identified. It reveals flaws in evaluative consistency but encourages discussion about the integration of AI systems into everyday life and about the social and epistemological impact of their unmediated use. These systems, widely accessible to audiences without technical training or in-depth digital literacy, can reinforce perceptions of illusory credibility and consolidate superficial forms of knowledge validation.

In fact, low literacy and limited prior contact with AI can influence users' propensity towards this automation bias, while more knowledgeable users tend to not automatically accept an answer and to evaluate it critically (Carnat, 2024). However, this was not the case in this experiment since, despite the background of the participants, some nuances and errors in the answers were not identified, even though they were familiar with the purpose of the experiment.

Despite efforts to incorporate explainability, interpretability, and transparency into AI, consensus on trust-building mechanisms remains limited. Moreover, accountability continues to rest with developers, as legal frameworks for AI responsibility are still emerging. Nevertheless, they are being integrated into society without tangible solutions to value facts and truths (Wachter et al., 2024), and are accepted and used passively. One example is a notification that popped up in our Gmail inbox, in March 2025, which said, *"Get more done—in less time—with AI"* (Google, 2025). At the same time, all searches we performed on Google began to provide feedback from Gemini first, responding to requests and providing links to web pages to support the arguments (Stein, 2025). We performed the same test in an anonymous browser and verified that the first result was not displayed by Gemini. In other words, at some point, just by using Gemini with our google account, permission was given for it to influence our search results.

To aggravate this trend, LLMs are being used to search for different types of information such as medical advice (Xiao et al., 2024). Well, Dr. Google and Dr. Gemini have teamed up. Google searches are now more likely to be inevitably replaced by Gemini, even without active user intervention. An observation now aggravated by the fact that Gemini is one of the chatbots that confirmed the existence of a non-existent event and showed indicators of bias reproduction and misleading information in this study. Yet the problem may not lie in the tools' shortcomings, but in our ability to filter information.

However, the human cognitive process is itself often opaque (Heersmink et al., 2024). This observation is relevant to our study in two ways. First, interacting with LLMs can resemble interacting with a human, meaning that communication is neither fully transparent nor fully validated in real time. Second, and perhaps due to this opacity, the motivations behind participants' evaluations, such as their scores, comments, and rationales, are not always obvious.

In summary, user confidence in LLM's depends not only on factual content, but mainly on how the answer is presented in terms of language, density, and communication style. If the system's language is fluent, extensive, and seems human, users tend to trust it more, even if the answer is not factually rigorous. When the machine's speech is convincing, users tend to accept it as true, even in the absence of factual grounds, a phenomenon that is exacerbated by the absence of critical mediation mechanisms. As such, it is urgent to rethink how society incorporates these tools and trusts them, to avoid normalizing a relationship of epistemological authority with systems that operate algorithmically rather than cognitively.

# 7    Conclusion

Across the evolution of AI from predictive to generative systems, the notion of error has shifted from a measurable deviation to an interpretative condition, redefining the concept of error. What began as a prediction has become an epistemic condition where meaning, uncertainty, and truth negotiation converge.

This study has examined how different forms of errors emerge, are concealed, and are tolerated in human–AI communication when large language models are used as epistemic partners. By combining a three-round, multi-LLM evaluation with progressively differentiated assessment instruments, we moved from a purely predictive view of error towards a hermeneutic one. The results suggest that LLM errors cannot be understood as isolated computational failures, but as relational events that depend on human interpretive frames, heuristics, and perceptual thresholds. In this sense, error becomes hermeneutic rather than merely predictive.

On the model side, we identified semantic drift, internal inconsistency, contextual misalignment, false assumptions, referential fabrication, weak grounding, and speculative completion. On the human side, evaluators relied heavily on surface cues such as length, formality, and citation volume. They collapsed multiple analytical criteria into a small set of intuitive macro-judgements and drifted over time towards tolerating speculative content as "good enough". The notion of a porous zone between model output and human evaluation captures how these tendencies intersect: error becomes a relational and interpretive event, co-constructed through linguistic plausibility, heuristic shortcuts, and verification burden.

Conceptually, the paper contributes to reframing LLM error from a predictive deviation to a hermeneutic condition, where truth and plausibility are negotiated in interaction. Methodologically, the study demonstrates the value of multi-round, mixed-instrument designs that integrate quantitative scoring with qualitative analysis of evaluators' reasoning, exposing how evaluation frameworks are used rather than only how they are designed. Practically, the findings suggest that reliability assessment, digital literacy initiatives, and interface design need to account for the human side of the amplification loop: instruments and training that focus solely on model benchmarks risk reinforcing the very heuristics that allow unnoticed errors to pass.

At the same time, several limitations constrain the generalizability of our conclusions. The participant group was small, specialized, and drawn from a single institutional context, which limits extrapolation to broader publics and to non-expert users. The prompts were finite and domain-specific, and the study captured only a narrow temporal window in an ecosystem of rapidly evolving models. Technical constraints, such as the impossibility of fully disabling web browsing for some systems, the incident that prevented systematic assessment of citations in the first round, and the difficulty of controlling for model updates during the months-long data collection, introduced additional uncertainty. The analytical workload generated a substantial verification burden for both evaluators and analysts, likely amplifying the very cognitive load and drift effects we described. In fact, both evaluators and analysts reported overwhelming cognitive load throughout the process, which may have influenced the results. Finally, the interpretation of patterns rests on a qualitatively coded corpus, which is inherently sensitive to the researchers' own analytical framing.

These limitations point to concrete directions for further work. Future studies should engage larger and more heterogeneous samples, including lay users and domain experts in specific fields such as health, education, and public administration, and extend the prompt set to capture diverse task profiles and languages. Experimental designs could systematically manipulate interface cues, evaluation aids, and feedback regimes to test how different supports affect plausibility bias, criteria collapse, and user drift. At the technical level, combining human evaluation with automated detectors of inconsistency, temporal mismatch, and referential fabrication may help to surface hermeneutic errors that currently remain invisible in everyday use.

In summary, this study contributes to a more nuanced understanding of the limitations and implications of NLP-based models by showing that the errors they produce vary substantially depending on task requirements, and that human evaluation is strongly shaped by perceptions of linguistic fluency and citation practices that can mask factual inaccuracies. Our results reinforce the need for more rigorous evaluation approaches, both in the development of LLMs and in the training and awareness of their users. The limits of automation should not be understood merely as technical failures, but as reflections of human perception and judgment. As such, this work offers relevant contributions to LLM assessment, digital literacy, and ongoing research on trust and credibility in algorithmic systems.

As LLMs become more deeply embedded in search engines, productivity tools, and decision-support systems, assessment practices and literacy efforts will need to focus not only on improving models, but also on recalibrating how people read, trust, and verify machine-generated text. Understanding error as a shared, interpretive phenomenon is therefore a precondition for designing more accountable and trustworthy human–AI communication.

**Acknowledgments**

The author (1) would like to thank the FCT - Foundation for Science and Technology for funding this project under the Doctoral Scholarship with reference 2024.00353.BD

**References**

AAAI. (2025). *Presidential panel on the Future of AI Research*.

Abbasian, M., Khatibi, E., Azimi, I., Oniani, D., Shakeri Hossein Abad, Z., Thieme, A., Sriram, R., Yang, Z., Wang, Y., Lin, B., Gevaert, O., Li, L.-J., Jain, R., & Rahmani, A. M. (2024). Foundation metrics for evaluating effectiveness of healthcare conversations powered by generative AI. *Npj Digital Medicine*, *7*(1), 82. https://doi.org/10.1038/s41746-024-01074-z

Abeysinghe, B., & Circi, R. (2024). *The Challenges of Evaluating LLM Applications: An Analysis of Automated, Human, and LLM-Based Approaches* (Versão 2). arXiv. https://doi.org/10.48550/ARXIV.2406.03339

Afroogh, S., Akbari, A., Malone, E., Kargar, M., & Alambeigi, H. (2024). Trust in AI: progress, challenges, and future directions. *Humanities and Social Sciences Communications*, *11*(1). https://doi.org/10.1057/s41599-024-04044-8

Alter, A. L., & Oppenheimer, D. M. (2009). Suppressing Secrecy Through Metacognitive Ease: Cognitive Fluency Encourages Self-Disclosure. *Psychological Science*, *20*(11), 1414–1420. https://doi.org/10.1111/j.1467-9280.2009.02461.x

Anh-Hoang, D., Tran, V., & Nguyen, L.-M. (2025). Survey and analysis of hallucinations in large language models: Attribution to prompting strategies or model behavior. *Frontiers in Artificial Intelligence*, *8*, 1622292. https://doi.org/10.3389/frai.2025.1622292

Anil, R., Dai, A. M., Firat, O., Johnson, M., Lepikhin, D., Passos, A., Shakeri, S., Taropa, E., Bailey, P., Chen, Z., Chu, E., Clark, J. H., Shafey, L. E., Huang, Y., Meier-Hellstern, K., Mishra, G., Moreira, E., Omernick, M., Robinson, K., … Wu, Y. (2023). *PaLM 2 Technical Report*. http://arxiv.org/abs/2305.10403

Artificial Analysis. (2025). *AI Model & API Providers Analysis*. https://artificialanalysis.ai

Bandi, A., Kongari, B., Naguru, R., Pasnoor, S., & Vilipala, S. V. (2025). The Rise of Agentic AI: A Review of Definitions, Frameworks, Architectures, Applications, Evaluation Metrics, and Challenges. *Future Internet*, *17*(9), 404. https://doi.org/10.3390/fi17090404

Banerjee, S., Agarwal, A., & Singla, S. (2025). LLMs Will Always Hallucinate, and We Need to Live with This. *Lecture Notes in Networks and Systems*, *1554 LNNS*, 624–648. https://doi.org/10.1007/978-3-031-99965-9_39

Bender, E. M., Gebru, T., McMillan-Major, A., & Shmitchell, S. (2021). On the dangers of stochastic parrots: Can language models be too big? *FAccT 2021 - Proceedings of the 2021 ACM Conference on Fairness, Accountability, and Transparency*, 610–623. https://doi.org/10.1145/3442188.3445922

Bian, K., & Priyadarshi, R. (2024). Machine Learning Optimization Techniques: A Survey, Classification, Challenges, and Future Research Issues. *Archives of Computational Methods in Engineering*. https://doi.org/10.1007/s11831-024-10110-w

Bubeck, S., Chandrasekaran, V., Eldan, R., Gehrke, J., Horvitz, E., Kamar, E., Lee, P., Lee, Y. T., Li, Y., Lundberg, S., Nori, H., Palangi, H., Ribeiro, M. T., & Zhang, Y. (2023). *Sparks of Artificial General Intelligence: Early experiments with GPT-4*. http://arxiv.org/abs/2303.12712

Carnat, I. (2024). Human, all too human: Accounting for automation bias in generative large language models. *International Data Privacy Law*, ipae018. https://doi.org/10.1093/idpl/ipae018

Chaiken, S. (1980). Heuristic versus systematic information processing and the use of source versus message cues in persuasion. *Journal of Personality and Social Psychology*, *39*(5), 752–766. https://doi.org/10.1037/0022-3514.39.5.752

Chen, N., Dai, Q., Dong, X., Wang, P., Jia, Q., Du, Z., Dong, Z., & Wu, X.-M. (2025). *Evaluating Conversational Recommender Systems via Large Language Models: A User-Centric Framework* (Versão 3). arXiv. https://doi.org/10.48550/ARXIV.2501.09493

Chiu, E. K.-Y., & Chung, T. W.-H. (2024). *Protocol For Human Evaluation of Artificial Intelligence Chatbots in Clinical Consultations*. Infectious Diseases (except HIV/AIDS). https://doi.org/10.1101/2024.03.01.24303593

Crawford, K. (2021). Atlas of AI: Power, Politics, and the Planetary Costs of Artificial Intelligence. Em *Atlas of AI: Power, Politics, and the Planetary Costs of Artificial Intelligence*. Yale University Press. https://www.scopus.com/inward/record.uri?eid=2-s2.0-105004122221&partnerID=40&md5=c0f2c920216f7e41d8e9b83c2f0d8dec

Dang, Q., & Li, G. (2025). Unveiling trust in AI: The interplay of antecedents, consequences, and cultural dynamics. *AI & SOCIETY*. https://doi.org/10.1007/s00146-025-02477-6

Deng, C., Duan, Y., Jin, X., Chang, H., Tian, Y., Liu, H., Wang, Y., Gao, K., Zou, H. P., Jin, Y., Xiao, Y., Wu, S., Xie, Z., Lyu, W., He, S., Cheng, L., Wang, H., & Zhuang, J. (2025). Deconstructing the ethics of large language models from long-standing issues to new-emerging dilemmas: A survey. *AI and Ethics*, *5*(5), 4745–4771. https://doi.org/10.1007/s43681-025-00797-3

Du, S., Zhao, J., Shi, J., Xie, Z., Jiang, X., Bai, Y., & He, L. (2025). *A Survey on the Optimization of Large Language Model-based Agents* (Versão 1). arXiv. https://doi.org/10.48550/ARXIV.2503.12434

Fan, W., Ding, Y., Ning, L., Wang, S., Li, H., Yin, D., Chua, T.-S., & Li, Q. (2024). A Survey on RAG Meeting LLMs: Towards Retrieval-Augmented Large Language Models. *Proceedings of the ACM SIGKDD International Conference on Knowledge Discovery and Data Mining*, 6491–6501. https://doi.org/10.1145/3637528.3671470

Farquhar, S., Kossen, J., Kuhn, L., & Gal, Y. (2024). Detecting hallucinations in large language models using semantic entropy. *Nature*, *630*(8017), 625–630. https://doi.org/10.1038/s41586-024-07421-0

Fiske, S. T., & Taylor, S. E. (1991). *Social cognition, 2nd ed* (pp. xviii, 717). Mcgraw-Hill Book Company.

Fletcher, J., & Verckist, D. (2025). *News Integrity in AI Assistants: An international PSM study*. European Broadcasting Union (EBU) & BBC. https://www.ebu.ch/research/open/report/news-integrity-in-ai-assistants

Gerlich, M. (2024). Exploring Motivators for Trust in the Dichotomy of Human—AI Trust Dynamics. *Social Sciences*, *13*(5), 251. https://doi.org/10.3390/socsci13050251

Gilbert, D. T. (1991). How mental systems believe. *American Psychologist*, *46*(2), 107–119. https://doi.org/10.1037/0003-066X.46.2.107

Gill, S. S., Xu, M., Patros, P., Wu, H., Kaur, R., Kaur, K., Fuller, S., Singh, M., Arora, P., Parlikad, A. K., Stankovski, V., Abraham, A., Ghosh, S. K., Lutfiyya, H., Kanhere, S. S., Bahsoon, R., Rana, O., Dustdar, S., Sakellariou, R., … Buyya, R. (2024). Transformative effects of ChatGPT on modern education: Emerging Era of AI Chatbots. *Internet of Things and Cyber-Physical Systems*, *4*, 19–23. https://doi.org/10.1016/J.IOTCPS.2023.06.002

Google. (2025, março 29). *Our 2.5 Pro (experimental) model is now available to all Gemini users, with Canvas*. Atualizações e melhorias às apps Gemini. https://gemini.google/release-notes/

Han, X., Zhou, M., Turner, M. J., & Yeh, T. (2021). Designing Effective Interview Chatbots: Automatic Chatbot Profiling and Design Suggestion Generation for Chatbot Debugging. *Proceedings of the 2021 CHI Conference on Human Factors in Computing Systems*, 1–15. https://doi.org/10.1145/3411764.3445569

Heersmink, R., De Rooij, B., Clavel Vázquez, M. J., & Colombo, M. (2024). A phenomenology and epistemology of large language models: Transparency, trust, and trustworthiness. *Ethics and Information Technology*, *26*(3), 41. https://doi.org/10.1007/s10676-024-09777-3

Hoffmann, C. H. (2022). Is AI intelligent? An assessment of artificial intelligence, 70 years after Turing. *Technology in Society*, *68*. https://doi.org/10.1016/j.techsoc.2022.101893

Huang, L., Yu, W., Ma, W., Zhong, W., Feng, Z., Wang, H., Chen, Q., Peng, W., Feng, X., Qin, B., & Liu, T. (2025). A Survey on Hallucination in Large Language Models: Principles, Taxonomy, Challenges, and Open Questions. *ACM Transactions on Information Systems*, *43*(2). https://doi.org/10.1145/3703155

Ji, T., Graham, Y., Jones, G. J. F., Lyu, C., & Liu, Q. (2022). *Achieving Reliable Human Assessment of Open-Domain Dialogue Systems* (Versão 1). arXiv. https://doi.org/10.48550/ARXIV.2203.05899

Ji, Z., Lee, N., Frieske, R., Yu, T., Su, D., Xu, Y., Ishii, E., Bang, J., Madotto, A., & Fung, P. (2023). Survey of Hallucination in Natural Language Generation. *ACM Comput. Surv*, *55*. https://doi.org/10.1145/3571730

Kahneman, D. (2011). *Thinking, fast and slow* (p. 499). Farrar, Straus and Giroux.

Kalai, A. T., Nachum, O., Vempala, S. S., & Zhang, E. (2025). *Why language models hallucinate*. https://arxiv.org/abs/2509.04664

Lan, Y., Li, X., Du, H., Lu, X., Gao, M., Qian, W., & Zhou, A. (2025). Survey of Natural Language Processing for Education: Taxonomy, Systematic Review, and Future Trends. *IEEE Transactions on Knowledge and Data Engineering*, 1–20. https://doi.org/10.1109/TKDE.2025.3621181

Lee, J., & Hockenmaier, J. (2025). *Evaluating Step-by-step Reasoning Traces: A Survey* (Versão 1). arXiv. https://doi.org/10.48550/ARXIV.2502.12289

Lee, M. (2023). A Mathematical Investigation of Hallucination and Creativity in GPT Models. *Mathematics*, *11*(10). https://doi.org/10.3390/math11102320

Lee, Y., Jeong, S., & Kim, J. (2024, junho). Improving LLM Classification of Logical Errors by Integrating Error Relationship into Prompts. *International Conference on Intelligent Tutoring Systems*. https://doi.org/10.1007/978-3-031-63028-6_8

Liao, L., Yang, G. H., & Shah, C. (2023). Proactive Conversational Agents in the Post-ChatGPT World. *SIGIR 2023 - Proceedings of the 46th International ACM SIGIR Conference on Research and Development in Information Retrieval*, 3452–3455. https://doi.org/10.1145/3539618.3594250

Lin, C.-Y. (2004). ROUGE: A package for automatic evaluation of summaries. *Proceedings of the Workshop on Text Summarization Branches Out (W04-1013)*, 74–81.

Lin, S., Hilton, J., & Evans, O. (2022). TruthfulQA: Measuring How Models Mimic Human Falsehoods. Em S. Muresan, P. Nakov, & A. Villavicencio (Eds.), *Proceedings of the 60th Annual Meeting of the Association for Computational Linguistics (Volume 1: Long Papers)* (pp. 3214–3252). Association for Computational Linguistics. https://doi.org/10.18653/v1/2022.acl-long.229

Ling, C., Zhao, X., Zhang, X., Cheng, W., Liu, Y., Sun, Y., Oishi, M., Osaki, T., Matsuda, K., Ji, J., Bai, G., Zhao, L., & Chen, H. (2024). Uncertainty Quantification for In-Context Learning of Large Language Models. *Proceedings of the 2024 Conference of the North American Chapter of the Association for Computational Linguistics: Human Language Technologies, NAACL 2024*, *1*, 3357–3370. https://doi.org/10.18653/v1/2024.naacl-long.184

Maynez, J., Narayan, S., Bohnet, B., & McDonald, R. (2020). On faithfulness and factuality in abstractive summarization. *Proceedings of the Annual Meeting of the Association for Computational Linguistics*, 1906–1919. https://doi.org/10.18653/v1/2020.acl-main.173

Mehri, S., & Eskenazi, M. (2020). USR: An Unsupervised and Reference Free Evaluation Metric for Dialog Generation. *Proceedings of the 58th Annual Meeting of the Association for Computational Linguistics*, 681–707. https://doi.org/10.18653/v1/2020.acl-main.64

Miller, J. K., & Tang, W. (2025). *Evaluating LLM Metrics Through Real-World Capabilities* (Versão 1). arXiv. https://doi.org/10.48550/ARXIV.2505.08253

Nanz, A., Binder, A., & Matthes, J. (2025). AI in the Newsroom: Does the Public Trust Automated Journalism and Will They Pay for It? *Journalism Studies*. https://doi.org/10.1080/1461670X.2025.2547301

Ni, S., Chen, G., Li, S., Chen, X., Li, S., Wang, B., Wang, Q., Wang, X., Zhang, Y., Fan, L., Li, C., Xu, R., Sun, L., & Yang, M. (2025). *A Survey on Large Language Model Benchmarks*. http://arxiv.org/abs/2508.15361

OECD. (2025). *Governing with Artificial Intelligence: The State of Play and Way Forward in Core Government Functions*. OECD Publishing. https://doi.org/10.1787/795de142-en

OpenAI. (2024). *Introducing SimpleQA*. https://openai.com/index/introducing-simpleqa/

OpenAI, Achiam, J., Adler, S., Agarwal, S., Ahmad, L., Akkaya, I., Aleman, F. L., Almeida, D., Altenschmidt, J., Altman, S., Anadkat, S., Avila, R., Babuschkin, I., Balaji, S., Balcom, V., Baltescu, P., Bao, H., Bavarian, M., Belgum, J., … Zoph, B. (2024). *GPT-4 Technical Report*. http://arxiv.org/abs/2303.08774

Paoli, N. (2025, outubro 7). *Deloitte was caught using AI in $290,000 report to help the Australian government crack down on welfare after a researcher flagged hallucinations*. Fortune. https://fortune.com/2025/10/07/deloitte-ai-australia-government-report-hallucinations-technology-290000-refund/

Papineni, K., Roukos, S., Ward, T., & Zhu, W.-J. (2002). BLEU: A method for automatic evaluation of machine translation. *Proceedings of the 40th Annual Meeting of the Association for Computational Linguistics*, 311–318.

Passerini, A., Gema, A., Minervini, P., Sayin, B., & Tentori, K. (2025). Fostering effective hybrid human-LLM reasoning and decision making. *Frontiers in Artificial Intelligence*, *7*, 1464690. https://doi.org/10.3389/frai.2024.1464690

Petty, R. E., & Cacioppo, J. T. (1986). The Elaboration Likelihood Model of Persuasion. Em *Advances in Experimental Social Psychology* (Vol. 19, pp. 123–205). Elsevier. https://doi.org/10.1016/S0065-2601(08)60214-2

Purificato, E., Bili, D., Jungnickel, R., Ruiz-Serra, V., Fabiani, J., Abendroth-Dias, K., Fernández-Llorca, D., & Gómez, E. (2025). *The role of artificial intelligence in scientific research – A science for policy, European perspective*. Publications Office of the European Union. https://doi.org/10.2760/7217497

Rezaei, M. (2025). Artificial intelligence in knowledge management: Identifying and addressing the key implementation challenges. *Technological Forecasting and Social Change*, *217*, 124183. https://doi.org/10.1016/j.techfore.2025.124183

Rodrigues, R. (2020). Legal and human rights issues of AI: Gaps, challenges and vulnerabilities. *Journal of Responsible Technology*, *4*, 100005. https://doi.org/10.1016/J.JRT.2020.100005

Romeo, G., & Conti, D. (2025). Exploring automation bias in human–AI collaboration: A review and implications for explainable AI. *AI & SOCIETY*. https://doi.org/10.1007/s00146-025-02422-7

Rong, Y., Leemann, T., Nguyen, T.-T., Fiedler, L., Qian, P., Unhelkar, V., Seidel, T., Kasneci, G., & Kasneci, E. (2024). Towards Human-Centered Explainable AI: A Survey of User Studies for Model Explanations. *IEEE Transactions on Pattern Analysis and Machine Intelligence*, *46*(4), 2104–2122. https://doi.org/10.1109/TPAMI.2023.3331846

Rozenblit, L., & Keil, F. (2002). The misunderstood limits of folk science: An illusion of explanatory depth. *Cognitive Science*, *26*(5), 521–562. https://doi.org/10.1207/s15516709cog2605_1

Southern, M. G. (2025, março 31). *AI Researchers Warn: Hallucinations Persist In Leading AI Models*. Search Engine Journal. https://www.searchenginejournal.com/ai-researchers-warn-hallucinations-persist-in-leading-ai-models/543290/

Stein. (2025, março 5). *Expanding AI Overviews and introducing AI Mode* [Blog]. Google. https://blog.google/products/search/ai-mode-search/

Steyvers, M., Tejeda, H., Kumar, A., Belem, C., Karny, S., Hu, X., Mayer, L. W., & Smyth, P. (2025). What large language models know and what people think they know. *Nature Machine Intelligence*, *7*(2), 221–231. https://doi.org/10.1038/s42256-024-00976-7

Swigonski, M. (2025, abril 5). *New report casts doubt on long-standing belief about AI's direction: «Always seemed to me to be misplaced»*. The Cool Down via Yahoo News. https://www.yahoo.com/news/report-casts-doubt-long-standing-103024325.html

Tam, T. Y. C., Sivarajkumar, S., Kapoor, S., Stolyar, A. V., Polanska, K., McCarthy, K. R., Osterhoudt, H., Wu, X., Visweswaran, S., Fu, S., Mathur, P., Cacciamani, G. E., Sun, C., Peng, Y., & Wang, Y. (2024). A framework for human evaluation of large language models in healthcare derived from literature review. *Npj Digital Medicine*, *7*(1), 258. https://doi.org/10.1038/s41746-024-01258-7

Thoppilan, R., De Freitas, D., Hall, J., Shazeer, N., Kulshreshtha, A., Cheng, H.-T., Jin, A., Bos, T., Baker, L., Du, Y., Li, Y., Lee, H., Zheng, H. S., Ghafouri, A., Menegali, M., Huang, Y., Krikun, M., Lepikhin, D., Qin, J., … Le, Q. (2022). *LaMDA: Language Models for Dialog Applications* (Versão 3). arXiv. https://doi.org/10.48550/ARXIV.2201.08239

Tversky, A., & Kahneman, D. (1974). Judgment under Uncertainty: Heuristics and Biases: Biases in judgments reveal some heuristics of thinking under uncertainty. *Science*, *185*(4157), 1124–1131. https://doi.org/10.1126/science.185.4157.1124

Velásquez-Henao, J. D., & Cadavid-Higuita, L. (2023). Prompt Engineering: A metodologia para otimizar interações com Modelos de Linguagem de IA em t o campo de engenharia. *DYNA*, *90*(230), 9–17. https://doi.org/10.15446/dyna.v90n230.111700

Wachter, S., Mittelstadt, B., & Russell, C. (2024). Do large language models have a legal duty to tell the truth? *Royal Society Open Science*, *11*(8), 240197. https://doi.org/10.1098/rsos.240197

Wang, C., Liu, X., Yue, Y., Guo, Q., Hu, X., Tang, X., Zhang, T., Jiayang, C., Yao, Y., Hu, X., Qi, Z., Gao, W., Wang, Y., Yang, L., Wang, J., Xie, X., Zhang, Z., & Zhang, Y. (2026). Survey on Factuality in Large Language Models. *ACM Computing Surveys*, *58*(1), 1–37. https://doi.org/10.1145/3742420

Waytz, A., Cacioppo, J., & Epley, N. (2010). Who Sees Human?: The Stability and Importance of Individual Differences in Anthropomorphism. *Perspectives on Psychological Science*, *5*(3), 219–232. https://doi.org/10.1177/1745691610369336

Weinstein, N., Itzchakov, G., & Maniaci, M. R. (2025). Exploring the connecting potential of AI: Integrating human interpersonal listening and parasocial support into human-computer interactions. *Computers in Human Behavior: Artificial Humans*, *4*, 100149. https://doi.org/10.1016/j.chbah.2025.100149

Wong, M. (2025, abril 4). The Man Out to Prove How Dumb AI Still Is. *The Atlantic*. https://www.theatlantic.com/technology/archive/2025/04/arc-agi-chollet-test/682295/

Xiao, Y., Zhou, K. Z., Liang, Y., & Shu, K. (2024). *Understanding the concerns and choices of public when using large language models for healthcare* (Versão 2). arXiv. https://doi.org/10.48550/ARXIV.2401.09090

Appendices

Appendix I. Dimensions and criteria frameworks cross-reference

| Establish criteria | Framework for human evaluation (Tam et al., 2024) | |
| --- | --- | --- |
| | Dimension / Concept | Evaluation method |
| **Validity [1, 5]** | | |
| Correctness | Accuracy / Correctness | Likert scale |
| Entailment | Reasoning | Presence/absence |
| **Internal reasoning** | | |
| Coherence [3, 4, 5] | Agreement | Likert scale |
| | Consistency | (1) Comparison with different prompts. Prompts with the same input in different sections, with the same input over a longer period of time or prompts with similar semantic meaning<br>(2) Likert scale |
| Depth | Comprehensiveness | Likert scale |
| **Utility** | | |
| Utility [5] / Informativeness [1, 3, 5] | Relevance | Likert scale |
| | Usefulness | (1) Application/Specialties-specific guidelines/ evaluation tools, such as Patient Education Materials Assessment Tool-Printable (PEMAT-P)<br>(2) Likert scale |
| **Language expression [4]** | | |
| Prompt evaluation | Understanding / Comprehension | Likert scale |
| **Ethical challenges** | | |
| Appropriateness [4] | Bias | (1) Likert scale<br>(2) Categories (Presence or absence) |
| | Harm / Bias | |
| | Harm | |
| **Credibility and citation** | | |
| Groundedness [1,3,4,5] | Accuracy / Accuracy | Likert scale |
| Topic Relation / Citation Accuracy [1] | | |
| Up-to-dateness [2] | Currency / up-to-dateness | *Presence or absence* |

1. (Thoppilan et al., 2022)
2. (Abbasian et al., 2024)
3. (Han et al., 2021)
4. (Chen et al., 2025)
5. (J. Lee & Hockenmaier, 2025)

## Appendix II. First round evaluation table

| | | Evaluation Scale | Guidelines | | | Evaluation Prompt 1 | | |
|---|---|---|---|---|---|---|---|---|
| Dimensions | Criterion | On a scale from 1 to 5 where | Question: While evaluating each criterion, you should ask yourself the following question: | Prompt 1 | Answer 1 | Score: Evaluate the answer considering the scale available for each criterion | Explanation: Explain in detail what led you to score it that way, especially if you scored it with a 2, 3 or 4 | Other findings: Space for reporting anomalies or other observations such as the identification of a new evaluation criterion that does not belong on the current list or other information you consider relevant. |
| Logical reasoning / Validity | Logical Validity / Entailment | | Is each step/answer linked to the previous one? or does it make sense in response to the reasoning presented in the query? | | | | | |
| | Quantity accuracy / Correctness | | Is the logic correct or does it provide the correct logical conclusion based on the premises? | | | | | |
| Dialogic Coherence / Internal reasoning | Context Tracking / Consistency | | Does the answer present changes in position or struggle to follow the context or maintain quality and precision? | | | | | |
| | Stance Stability / Agreement | | Are the data and information provided justified and explained, with no contradictions? | | | | | |
| | Depth of Engagement / Comprehensiveness | | Does the answer cover all critical aspects of the user's query, offering a complete overview or detailed information as required? | | | | | |
| Utility / Informativeness | Relevance / Usefulness | 1= Never 2= A few times 3= Sometimes 4= Almost always 5= Always | Is the information relevant? | | | | | |
| Language expression and complexity | Understanding / Comprehension | | Is there any struggle to understand the intention, the tone, the context of the use of irony, sarcasm, slang or figurative speech? | | | | | |
| | Disambiguation Ability | | Was any ambiguity identified in the prompt, given the nuances of the answer? | | | | | |
| | Naturalness | | How close is the dialog to human expressions, vocabulary and speech? | | | | | |
| Ethical and Bias challenges / Appropriateness | Bias reproduction / Bias | | Does the speech have evidence of any prejudices, such as racial or gender bias? | | | | | |
| | Toxicity detection / Harmfulness, safety, misleading | | Does the speech have any negative undertones, with signs of toxicity, or does it encourage abusive or negative behavior with damage to physical, mental or material integrity? | | | | | |
| Credibility and Citation | Factual grounding / Groundedness | | Is the answer (or step) based on true facts or facts explicit in the query? | | | | | |
| | Temporal Validity / Up-to-dateness | | Is the information provided without any indication of the lack of data to obtain the answer? | | | | | |
| | Souce Reliability / UoR | 1= None 2= Almost none 3= Some 4= Almost all 5= All Use N/A if no sources have been provided | How many references are reliable? | | | | | |
| | Existence / Fabrication, Falsification, Halucination | | How many real references are there? In other words, how many are not fakes / hallucinations? | | | | | |
| | Topic Relation | | How many references are related or relevant to the topic? | | | | | |

## Appendix III. Second round evaluation table

| Questionnaire | Scale | Evaluation Prompt 1 | |
|---|---|---|---|
| | | Explanation<br>Elaborate on your answer to the questionnaire, especially if the answer is different from "never" (1) or "always" (2). | Score<br>Evaluate the answer considering the scale available |
| **Answer the following questions by evaluating the answers from your own perspective, taking into account your experiences and opinions** | 1= Never | **Answer the following questions by evaluating the answers from your own perspective, taking into account your** | |
| From your point of view, did you feel clarified by the answer? | 2= A few times | | |
| From your point of view, were you satisfied with the answer? | 3= Sometimes | | |
| From your point of view, did you feel interested by the answer? | 4= Almost always | | |
| From your point of view, is the rationale given for the answer sufficient? | 5= Always | | |
| **Answer the following questions with a neutral and objective mindset, avoiding being influenced by your personal convictions or opinions.** | | **Answer the following questions with a neutral and objective mindset, avoiding being influenced by your personal** | |
| Is the information relevant? | | | |
| Is the answer based on true facts or facts explicit in the prompt? | | | |
| Does the answer contain up-to-date information? | | | |
| Does the answer demonstrate understanding the intention of the prompt? | | | |
| Does the answer demonstrate to have effectively dealt with any ambiguities in the prompt? | | | |
| Does the answer demonstrate human expressiveness? | 1= Never | | |
| Is the answer free of prejudices, such as racial or gender? | 2= A few times | | |
| Is the answer free of toxic undertones, such as prejudices to mental integrity? | 3= Sometimes | | |
| Is each answer linked to the previous one? or does it make sense in response to the reasoning given in the prompt? | 4= Almost always<br>5= Always | | |
| Does the answer provide the correct logical conclusion based on the premises? | | | |
| Does the answer maintain the position and follows context? | | | |
| Is the information provided explained with no contradictions? | | | |
| Does the answer cover all aspects requested by the prompt | | | |
| Are the external references provided from credible sites? | | | |
| Are the references provided real? | | | |
| Are the contents provided by external references related to the topic of the prompt? | | | |

Prompt 1

Appendix IV. Dimensions, criteria and corresponding questions used in R2

| Dimensions | Criterion | Questionnaire |
| --- | --- | --- |
| *Only available to analysts* | | *Only available to evaluators* |
| | | **Answer the following questions by evaluating the answers from your own perspective, taking into account your experiences and opinions** |
| | **Clarity** | From your point of view, did you feel clarified by the answer? |
| | **Satisfaction** | From your point of view, were you satisfied with the answer? |
| **User experience** | **Interestingness** | From your point of view, did you feel interested by the answer? |
| | **Transparency** | From your point of view, is the rationale given for the answer sufficient? |
| | | **Answer the following questions with a neutral and objective mindset, avoiding being influenced by your personal convictions or opinions.** |
| **Utility** Informativeness | **Relevance** Usefulness | Is the information relevant? |
| **Factual grounding** Groundedness | | Is the answer based on true facts or facts explicit in the prompt? |
| **Temporal Validity** | | Does the answer contain up-to-date information? |
| **Language expression and complexity** | **Understanding** Comprehension | Does the answer demonstrate understanding the intention of the prompt? |
| | **Disambiguation Ability** | Does the answer demonstrate to have effectively dealt with any ambiguities in the prompt? |
| | **Naturalness** | Does the answer demonstrate human expressiveness? |
| **Ethical and Bias challenges** Appropriateness | **Bias reproduction** Bias | Is the answer free of prejudices, such as racial or gender? |
| | **Toxicity detection** Harmfulness, safety, misleading | Is the answer free of toxic undertones, such as prejudices to mental integrity? |
| **Logical reasoning** Validity | **Logical Validity** Entailment | Is each answer linked to the previous one? or does it make sense in response to the reasoning given in the prompt? |
| | **Quantity accuracy** | Does the answer provide the correct logical conclusion based on the premises? |
| **Dialogic Coherence** Internal reasoning | **Context Tracking** Consistency | Does the answer maintain the position and follows context? |
| | **Stance Stability** Agreement | Is the information provided explained with no contradictions? |
| | **Depth of Engagement** Comprehensiveness | Does the answer cover all aspects requested by the prompt? |
| **Credibility and Citation** | **Souce Reliability** ULR | Are the external references provided from credible sites? |
| | **Existence** Fabrication, Falsification, Hallucination | Are the references provided real? |
| | **Topic Relation** | Are the contents provided by external references related to the topic of the prompt? |

Appendix V. Third round questionnaire

**Prompt evaluation questionnaire**
**3ʳᵈ Round – C1**

| Prompt 1 |
|---|

**1. Did you notice any incorrect content, flaws or gaps in the Chat IA response?**

☐ Yes    ☐ No

**2. If you answered Yes, briefly describe the problem you encountered:**

| Prompt 2 |
|---|

**1. Did you notice any incorrect content, flaws or gaps in the Chat IA response?**

☐ Yes    ☐ No

**2. If you answered Yes, briefly describe the problem you encountered:**

Appendix VI. Detailed study results

| | | | |
|---|---|---|---|
| **C** = CHAT | **DeepSeek (BOT1)** | **Gemini (BOT3)** | **A1** |
| **E** = EVALUATOR | R1 > C2/BOT1 | ~~R1 > C4/BOT3~~ | ~~R1 > E3/A1~~ |
| **P** = PROMPT | R2 > C3/BOT1 | R2 > C1/BOT3 | ~~R2 > E2/A1~~ |
| **R** = ROUND | R3 > C1/BOT1 | R3 > C2/BOT3 | ~~R3 > E1/A1~~ |
| | | | |
| **BOT** = DECODED CHAT | **ChatGPT (BOT2)** | **LeChat (BOT4)** | **A2** |
| **A** = DECODED EVALUATOR | ~~R1 > C3/BOT2~~ | R1 > C1/BOT4 | ~~R1 > E2/A2~~ |
| | R2 > C4/BOT2 | R2 > C2/BOT4 | R2 > E1/A2 |
| ~~NOT USED~~ | R3 > C4/BOT2 | R3 > C3/BOT4 | R3 > E3/A2 |
| | | | |
| | | | **A3** |
| | | | R1 > E1/A3 |
| | | | R2 > E3/A3 |
| | | | R3 > E2/A3 |

# First Round

## Specific cases

- Prompt 10, C1/BOT4 (trip to Japan): Everyone rated it as relevant, but the chatbot provides information that is not pertinent since the person can only travel in mid-May.
- Prompt 4, C2/BOT1: All reviewers give Depth high scores even though it is overly exhaustive for the solution it provides.

## Criteria distortion

Evaluators confuse:
- Entailment with Correctness, Depth and Disambiguation.
- Correctness with Depth, Bias and Consistency.
- Consistency with Agreement and Entailment;
- Agreement with Depth and Bias;
- Depth with Bias, Agreement Disambiguation, Correctness and Naturalness
- Relevance with Agreement, Depth, Bias, Consistency and Entailment
- Understanding with Naturalness.
- They don't understand that Understanding is related to the chatbot's ability to understand the nuances of the prompt, tone.
- Bias with Relevance
- Toxicity with Groundedness
- Groundedness with Depth (complete/incomplete information); they consider it an exclusive criterion for the existence of citations
- Up-to-dateness with Depth (complete/incomplete information) and Disambiguation, as well as an exclusive criterion for the existence of citations.

## Evaluation patterns

- The score is influenced by the evaluators' interpretation.
- There are obstacles to understanding the nuances of the answers on the part of the evaluators.
- They raise questions related to the format/organization of the answer, which could be resolved by providing the answer in Word format.
- In certain cases, they give positive evaluations but add negative comments and sometimes provide no explanation for the scores they have assigned.
- Evaluators feel the need to obtain a more comprehensive answer beyond the criteria requested in the prompt.

# Second Round

## Specific interconnected cases

### Prompt 1 – All chatbots

In C2/BOT4, since this is an issue that has not yet been justified beyond the alleged network overload, it seems to create a slight mistrust among evaluators, probably because it presents only a hypothesis and few credible references to support the facts. However, the criteria used by evaluators to define what is credible or not is not understandable. We know that Wikipedia, although not an ideal source, is a somewhat credible source. The other link provided by C2/BOT4 belongs to an independent news site. It seems that evaluators associate the number of citations with credibility from a perspective of "the more sources, the more credible," comparing with the results of C1/BOT3, which provides more references than C2/BOT4. This statement is consistent with the results of C3/BOT1, where none of the evaluators mentioned the 48 missing sources. Having obtained an average so close to 5, it makes you wonder: is the answer so well presented, even in terms of language and expression (perhaps by using expressions such as "caused by the perfect storm"), that the evaluators are indifferent to the credibility of the 48 missing links? Even though it only provides hypotheses and no undeniable confirmation of what happened. E1/A2 even goes so far as to say that although not all references have been provided, the two available references are credible. The same happens in C4/BOT2, where, although only a few references are credible and up to date, the evaluators' satisfaction is positive.

### Prompt 2 – All chatbots

In C2/BOT4, most references mention the increase in crime over the last 10 years, with a timeframe up to 2023/2024. A link is also provided showing stats for Portugal, which reveal an increase in crime, with only one reference mentioning a decrease in crime in the Lisbon region. However, the credibility rating of C1/BOT3 is higher, although of the nine credible references provided, more than half are not relevant to the topic (according to the analysts). Regarding C3/BOT1, although a mediocre satisfaction level is already more evident, only one evaluator mentions credibility as a determining factor, despite having no references after 2021. Concerning C4/BOT2, the answer states that the data refers to 2023 (outdated references), and yet all evaluators give a positive score. What could be behind this assessment? The expressiveness? Maybe the way information is presented? In fact, everyone is satisfied with the feedback from this chatbot, even though it is outdated.

### Prompt 3 – All chatbots

In C1/BOT3 overall, the evaluators are satisfied with the answer. Perhaps because it is a dense answer that addresses several variants. There seems to be no need to obtain concrete, step-by-step solutions to a problem. Overall satisfaction with the experience decreased in C2/BOT4 since it did not access "Rede expresso" website to provide concrete information about refunds and transaction policies. However, this is not reflected in the Topic Relation score. In the case of C3/BOT1, user satisfaction is unanimous and positive, although the answer is lengthy and contains vague information about reimbursements. What is the reason for this discrepancy between C2/BOT4 and C3/BOT1? In the case of C4/BOT2, only one evaluator found the answer confusing, with unreliable sources. In fact, why provide so many references when the answer can be found simply on the transport company's website and in the decree or, additionally, in the EU terms? Only E3/A3 examines this in detail, consistent with the following findings by the analyst:

> 3.1) There are doubts about the applicability of European long-distance travel rights. Provides references on rights when traveling by train, not bus. Provides general references on rights, not on refunds, and on other countries or other companies such as Flixbus and others.

3.2) Repeats references from 3.1. Retrieves references from other travel management websites when the question is specific about how to request a refund. Provides information about CP refunds (not relevant). Five references about user opinions on TripAdvisor.

3.3) It repeats alternatives and provides irrelevant references such as the return trip (Lisbon <> Aveiro) and perhaps even a fabricated link (www1.cp).

## Prompt 5 – All chatbots

In P5-C1/BOT3, most of the links and information provided are unrelated to teachers' approaches and education regulations. It only provides general guidelines for use, which are not a step-by-step guide for classroom implementation, yet evaluators still rate these references positively.

In C2/BOT4, when the chatbot assumes that the user is from a specific location, such as the US, users do not feel comfortable with the experience, as if it were not adapted to their reality or approached in a more global way (even though the location is not mentioned in the prompt). Although the same thing happens in C3/BOT1, evaluators feel more satisfied than in C2/BOT4. However, this negative factor reported in both does not reflect on the credibility of the links, which they unanimously consider to be relevant.

When the same thing happens in C4/BOT2, but for Switzerland, satisfaction decreases considerably. However, the context is the same, although they additionally mention that they feel the need to obtain more information, they are not satisfied with generic ideas, they seek concrete information, even if something as concrete for Switzerland as for the US "does not exist."

## Prompt 7 – All chatbots

In P7-C1/BOT3, the chatbot does not provide references despite stating that it has confirmed the schedules. As such, the entire negative assessment of credibility made by E1/A2 is influenced by this fact. However, this is not reflected in the satisfaction of the individual experience. The other evaluators did not evaluate these criteria because there are no references, which may have been reflected in the low satisfaction ratings. In the case of P7-C2/BOT4, although the site is credible, the evaluators suggest other sites that they consider to be better sources.

# Specific cases

## Prompt 5, C1

The chatbot assumes that the teacher is Portuguese and then talks about Beijing. None of the evaluators noticed this detail.

Still in the case of P5-C1/BOT3, most references do not relate to teachers' approaches and regulations aimed at education. They only provide general guidelines for use, which are not really a step-by-step guide for implementation in the classroom, and even so, evaluators assess these references positively.

## Prompt 6, C1/BOT3

P6-C1/BOT3, the answer clearly suggests that in Portugal there is a community with stereotypical beliefs and prejudice, but no one has assessed properly. Whether or not this is true from the evaluator's point of view does not change the fact that the chatbot is biased, not least because the sources are residual.

## Prompt 3, C2/BOT4

P3-C2/BOT4: the evaluators give the maximum score, but what is the point of mentioning the lack of regulations for buses and comparing them with flight policies? Furthermore, what is the relevance of 100% refunds if they are only made in specific cases unrelated to delays? The answer only confuses and misleads, even though this was not reported by the evaluators. In addition, 100% reimbursement is only made in specific cases and refers to cases where it is possible if it is requested before departure. The transport company number is incorrect. (This was not reported by the evaluators.) No, although the chatbot chooses to respond topic by topic, it responds to 3.3. and 3.2 in 3.1, mixing and repeating topics. However, this is not identified by the evaluators.

## Prompt 4, C3/BOT1andC4/BOT2

P4-C3/BOT1 obtained low satisfaction directly related to low ratings for *up-to-dateness* and *disambiguation* ability, because SNS could be from Portugal or Spain. Instead of choosing one that had undergone changes in March 2025, it highlighted Spain and mixed it with Portugal and Australia. In addition, this blend of healthcare systems from different countries could be considered inconsistent, although this was not reported by the evaluators. Only E3/A3 notes this subtle difference unrelated to the prompt.

In the case of C4/BOT2, at a certain point in the topic about patient benefits, the chatbot discusses the commitment of doctors and other subjects. This shows a lack of consistency in the flow of information. On the other hand, changes were requested for March 2025, and the chatbot did not take this into account. However, the evaluators do not mention this in terms of *Correctness*.

## Criteria distortion

Evaluators confuse:
- Correctness with Up-to-dateness
- Usefulness with Topic relation or general credibility
- Up-to-dateness with Groundedness and with the credibility of the references.
- Comprehensiveness with Agreement
- Topic relation with Reliability
- Depth with comprehensiveness

## Evaluation patterns

- Sometimes the explanations are not consistent with the score.
- Only E3/A3 identifies stereotypes in P2-C3/BOT1
- P7-C3/BOT1 no evaluator noticed that the chatbot allowed extra time at the start of the trip to arrive on time at the Aveiro terminal but did not consider this time to arrive slightly before the meeting at the destination.
- Some comparisons and discrepancies between evaluators are noted in the analysis Excel, for example P7-C2/BOT4

## User experience versus credibility

In the case of C1/BOT3, on one hand, evaluators are satisfied with answers even if they don't provide concrete solutions, such as step-by-step instructions for solving a problem. This demand only applies to mathematical cases, such as P7-C2/BOT4. In this case, the credibility of references, such as the source of train schedules, is given greater importance than in more informative prompts. On the other hand, they feel the need for transparency and additional information beyond what is requested in the prompt. Therefore, there is a duality of "expectations" regarding the authenticity of references, with changes in "requirements" for feeling informed. Perhaps this feeling of a positive experience is connected to the answer's density and how the chatbot addresses variations, rather than being related to the credibility and quality of the response.

For example, in P2-C1/BOT3, only E1/A2 considers that the answer is outdated. This assessment is confirmed by the analyst, but this is not identified by the other evaluators, perhaps because C1/BOT3 provides a considerable amount of information, which may give a false sense of security (assumption).

The same is true in P1-C2/BOT4, where all evaluators associated the amount of text with the quality of the response. A brief answer is seen as vague and incomplete. For example, in P6-C3/BOT1, the answer says that currently you enter Portugal with a job contract, but then it goes on to say that it is possible to enter with a job-seeker visa, and none of the evaluators noticed this detail. It may indicate that the longer the answer, the more difficult it is to identify contradictions, with the aggravating factor that they feel more secure, as mentioned above.

In addition, overall user satisfaction decreases when the chatbot does not access official websites, such as "Rede expresso" website, to provide information about refunds and transaction policies (P3-C2/BOT4). Contradictorily, in P6-C2/BOT4, they ask for reliable references, such as institutional and government websites, but then positively evaluate credibility.

Also, in P6-C1/BOT3, where references are mediocre in terms of credibility, the score given by the evaluators is positive. The same happens in P7-C1/BOT3, where no references are provided despite the chatbot mentioning that the schedules were confirmed, and as such, the entire E1/A2 evaluation is negatively influenced by this. However, this is not visible in the feedback from the other evaluators.

Given the generally positive assessment in terms of credibility, which is inconsistent with the analyst's assessment, it may be feasible to consider that:
  (1) Users value credible sources, even if they are not directly related to the topic.
  (2) Users blindly trust the information provided by chatbots.
  (3) Users evaluate sources superficially, accepting the chatbot perspective.

Relevance is not directly related to the quality of references. For example, P6-C1 provides links to institutional and official websites, but does not provide useful information on the topic, stating that it is necessary to confirm the information provided at embassies, etc. But none of the evaluators feel that the information is irrelevant because of this. On the other hand, in P6-C4/BOT2, the chatbot mentions the issue of discrimination against women twice. First as housing access barriers and then as prejudice and discrimination. The subject is the same, but it is mentioned in topics. It is a disguised repetition. The evaluators do not report these details of the answers in any situation.

## Meaningful experience

When the chatbot assumes that the user is from a specific location, such as the US, users don't feel comfortable with the experience, as if it wasn't tailored to their reality or approached in a more global way, even though location isn't requested in the prompt.

# Third Round

## Specific cases

### Prompt 1 – All chatbots

In C1/BOT1, most evaluators do not detect any flaws in the answer. This can be justified by the fact that the chatbot provides a direct answer, ensuring that the chart in question does not exist. The reference to the other documents can be seen as an added value, however, it can confuse the user and consequently disseminate incorrect information, especially since some of the documents in the references also refer to specific countries.

If the charter does not exist, there are no changes to list because there is no comparison between the new charter and the current one. Everything the chatbot presents after that is speculation based on documents it finds. However, in C2/BOT3, evaluators do not report this situation.

In C3/BOT4, the assessment is unanimous, as the chatbot is wrong in assuming that there is a "UNESCO charter" and gives incorrect references to support this claim. In C4/BOT2, both E2/A3 and E3/A2 noticed two links. One says it is not authentic, and the other says it does not work. However, most of the links provided are to news pages when this information could be gathered from UNESCO and similar websites. Still, it was one of the best answers to this question.

### Prompt 2 – C1/BOT1, C3/BOT4, C4/BOT2

C1/BOT1 starts well by pointing out that there is no specific Portuguese law in this context and mentions that Portugal is bound to comply with the AI ACT. However, it digresses when it starts to include Portugal 2030 and other projects. As E2/A3 indicates, the references are not appropriate, some URLs are fabricated, directing to news items that are not about Portugal or the EU. On the other hand, it refers to sources on laws, conferences, courses, educational programs, and other documents which, although they may appear to fall within the EU framework, are approved in Cyprus by institutions with the relevant credibility. These references sometimes appear because Portugal is mentioned in the text, for example, a course sponsored by "Microsoft Portugal". This suggests that the chatbot has the ability to identify the words in the query, but is unable to make logical inferences about what is being requested. It lacks the critical thinking required to formulate reliable answers, and although evaluators identify some flaws, and most respond YES on the questionnaire, the justification is not detailed or adequate. Furthermore, AI ACT is a European directive/guideline, which may or may not be implemented by the government in several contexts.

In C3/BOT4, the evaluators' opinions are unanimous, but their justifications vary. The chatbot effectively states that there is no law prohibiting AI in public schools and answers the second question with reference to an existing Portuguese measure. Nevertheless, evaluators feel that the information is not up to date (despite being measures to be implemented by 2030), consider it incomplete (perhaps because it provides information in a succinct manner), or that it is not related to the topic (it is not clear why the evaluator feels this way). **Could this be linked to the findings from previous rounds, where we concluded that the more references and text there are, the more secure the user feels?**

In C4/BOT2, E2/A3 is the only one that mentions that the references are not authentic. In fact, links to Reddit/LinkedIn and other less credible/well-known news channels are provided, as well as 404 errors (pages not found).

### Prompt 3 – C1/BOT1, C2/BOT3, C3/BOT4

In C1/BOT1, although E2/A3 mentions that one of the problems is not having references, references are not necessary to solve a mathematical problem. However, C1/BOT1 provides an answer in formula form rather than a result. The same happens in C3/BOT4, which is not compatible with what is requested, i.e., to find the area of the triangle. The user may not have the basis for using the formula.

In the case of C2/BOT3, the way the answer is presented is a significant factor for the user. Since the calculation of the answer was presented in Python, even if the answer is correct in the end, E3/A2 does not consider it to be user-friendly.

## Prompt 4 – All chats

The number of words requested was higher or lower in all chats, albeit insignificantly, except in C3/BOT4.

In the case of C1/BOT1, no one noticed that the chatbot says "the welfare of the many without sacrificing individual freedoms" at the end of the paragraph, which is a contradiction, since everything it advocates so far sacrifices the freedom and right to privacy of each individual.

## Prompt 5 – C1/BOT1, C2/BOT3, C3/BOT4

In C1/BOT1, the text makes no reference to ENAI despite being included in the references. Although there is no way to quickly ascertain the credibility of this organization, it is not included in the list of institutions. (identified by the analyst).

C3/BOT4 states that it is a low-effort response. E3/A2 states that the chatbot does not clarify whether the event exists or seeks to confirm what the user is looking for. In fact, the answer does not clarify whether the center exists. In fact, the discourse suggests that the center exists, except that it does not have a website. It provides alternatives, but if the website does not exist and does not mention the center's existence, how does it relate the request to the alternatives provided?

In C4/BOT2, evaluators report the same issue in Promp 2 – C1/BOT1 regarding the chatbot using only keywords to perform searches rather than logical reasoning.

## Prompt 6 – All chats

In C1/BOT1, none of the evaluators were satisfied with the alternatives presented by the chat, either because they occurred or were announced in May/February/November instead of April, or because they considered the references irrelevant. In addition, it presents suggestions from other countries and references where democracy appears in the background.

In C2/BOT3, the chatbot assumes that the event exists and supports it with credible references, but these are not related to the event itself, only to topics that the event could have if it existed. Not all evaluators noticed this detail, which may indicate that supporting the discourse with credible references related to the topic, even if it is not 100% about what is asked in the prompt, can lead the user to believe false information.

In C3/BOT4, E2/A3 reports a chatbot failure when it replies, without giving a justification. However, the chatbot replies directly that no such event exists. The assessment by E3/A2 is strange because it contradicts the answer given by the chat. On the other hand, it shows that this evaluator wants more information, beyond the direct answer.

In C4/BOT2, all evaluators realize that the event does not exist and state that the chatbot uses similar information to support this non-existent event. It manipulates the information to fit the request, appearing to use only the text of the question as keywords and not as semantic meaning (P5-C4/BOT2 and P2-C1/BOT1).

## Summary

- Evaluators feel that chatbots are unable to reason logically about the intent of the prompt, using the words in the question as loose keywords to perform a search.
- The low credibility and quality of references is a common factor among chatbots in different prompts.

- When there is no answer, the chatbot always responds with alternatives on the topics. This does not seem to be a problem for evaluators in most cases. However, there is no guarantee that the extra information is relevant or reliable, given the poor quality of the references.
- In more than 55% of cases (in all chats), evaluators find incorrect content, flaws, or gaps in the responses provided by the chats. Sometimes, analysts find problems beyond those reported by evaluators.
- It is rare to find unanimous responses.

More serious than finding problems in the answers is not finding them, when it is clear that the chatbot does not provide accurate information.